\documentclass[]{aa}

\usepackage{times}
\usepackage{graphics}
\usepackage{epsf}
\begin{document}

\thesaurus{02    
(
11.05.2;    
11.19.3;    
12.03.3;    
13.09.1     
) }

\title{Far infrared extragalactic background radiation: 
I. Source counts with ISOPHOT }

\titlerunning{Far Infrared Extragalactic Background}

\author{
M.~Juvela \inst{1}
\and K.~Mattila \inst{2}
\and D.~Lemke \inst{3}
}

\offprints{M.~Juvela}

\institute{
$^1$Helsinki University Observatory, T\"ahtitorninm\"aki, P.O.Box 14, 
SF-00014 University of Helsinki, Finland (mjuvela@astro.helsinki.fi) \\
$^2$Helsinki University Observatory, T\"ahtitorninm\"aki, P.O.Box 14, 
SF-00014 University of Helsinki, Finland \\
$^3$Max-Planck-Institut f\"ur Astronomie, K\"onigstuhl 17, D-69117 Heidelberg, Germany
}

\date{Received <date> ; accepted <date>}

\maketitle

\begin{abstract}

As a part of the ISOPHOT CIRB (Cosmic Infrared Background Radiation) project
we have searched for point-like sources in eight fields mapped at two or
three wavelengths between 90$\mu$m and 180$\mu$m. Most of the 55 sources
detected are suspected to be extragalactic and cannot be associated with
previously known objects. It is probable, also from the far-infrared (FIR)
spectral energy distributions, that dust-enshrouded, distant galaxies form a
significant fraction of the sources.

We present a tentative list of new extragalactic FIR-sources and discuss the
uncertainties involved in the process of extracting point sources from the
ISOPHOT maps. Based on the analyzed data we estimate the number density of
extragalactic sources at wavelengths 90$\mu$m, 150$\mu$m and 180$\mu$m and
at flux density levels down to 100\,mJy to be 1$\cdot$10$^5$\,sr$^{-1}$,
2$\cdot$10$^5$\,sr$^{-1}$, and 3$\cdot$10$^5$\,sr$^{-1}$, respectively.

Strong galaxy evolution models are in best agreement with our
results, although the number of detections exceeds most model predictions.
No-evolution models can be rejected at a high confidence level.

Comparison with COBE results indicates that the detected sources
correspond to $\ga$20\% of the extragalactic background light at 90$\mu$m.
At longer wavelengths the corresponding fraction is $\sim$10\%.

\end{abstract}

\section{Introduction}

The cosmic infrared background (CIRB) consists in the far-infrared of the
integrated light of all galaxies along the line of sight plus any
contributions by intergalactic gas and dust, photon-photon interactions
($\gamma$-ray vs. CMB) and by hypothetical decaying relic particles. A large
fraction of the energy released in the universe since the recombination
epoch is expected to be contained in the CIRB. An important aspect is the
balance between the UV-optical and the infrared backgrounds: what is lost by
dust obscuration will re-appear through dust emission in the CIRB. Some
central, but still largely open, astrophysical problems to be addressed
through CIRB measurements include the formation and early evolution of
galaxies, and the star formation history of the universe.

The primary goal of the ISOPHOT CIRB project is the determination of the
flux level of the FIR CIRB. The other goals are the measurement of its
spatial fluctuations and the detection of the bright end of FIR point source
population contributing to the CIRB. 
The full analysis of the data from the DIRBE (Hauser et al.
\cite{hauser}; Schlegel et al.\cite{schlegel98}) and FIRAS (Fixsen et al.
\cite{fixsen98}) experiments indicated a CIRB at a surprisingly
high level of $\sim$1 MJy\,sr$^{-1}$ between 100 and 240 $\mu$m. Preliminary
results had been obtained already by Puget et al. (\cite{puget}). Lagache et
al. (\cite{lagache99a}) detected a component of Galactic dust emission
associated with warm ionized medium and the removal of this component lead
to a CIRB level of 0.7\,MJy\,sr$^{-1}$ at 140$\mu$m.

Because of the great importance of the FIR CIRB for cosmology these results
definitely require confirmation by independent measurements. 
ISOPHOT observation technique is different from COBE: (1) with relatively
small f.o.v. ISOPHOT is capable of looking at the darkest spots between the
cirrus clouds; (2) ISOPHOT has good sensitivity in the important FIR window
at 120 -- 200 $\mu$m; (3) with the good spatial and spectral sampling
ISOPHOT gives the possibility of recognizing and eliminating the emission of
galactic cirrus.

In the ISOPHOT CIRB project we have mapped four low-cirrus regions at high
galactic latitude at the wavelengths of 90, 150, and 180 $\mu$m (see
Fig.~\ref{fig:allsky}). Through this multi-wavelength mapping we will try to
separate the cirrus component and confirm the detection of sources at
neighboring wavelengths. In addition, we have performed absolute photometry
in several filters between 3.6 -- 200 $\mu$m at the darkest spots of the
fields. This photometry will be used (1) to secure the zero point for the
maps at 90, 150, and 180 $\mu$m, and (2) to determine the contribution by
the zodiacal emission using measurements of its SED at mid-IR wavelengths
where it dominates the sky brightness.

This paper presents the first step in the analysis of the ISOPHOT CIRB
observations. Here we will concentrate on the data reduction and the study
of the point sources (galaxies) found in the FIR maps.
The source counts determined in the FIR are important for the study of the
star formation history of the universe and for the testing of the current
models of galaxy evolution.

With recent observations at infrared and sub-mm wavelengths it has become
obvious that star formation efficiencies derived from optical and UV
observations only (e.g. Madau et al. \cite{madau}; Steidel et al.
\cite{steidel}, \cite{steidel99}; Cowie et al.
\cite{cowie96}, \cite{cowie97}, \cite{cowie98}; Hu et al. \cite{hu})
underestimate the true star formation activity at high redshifts
because the correction for dust extinction is unknown (e.g. Heckman et al.
\cite{heckman}).

IRAS has shown that in the local universe about one third of the luminosity
is emitted at infrared wavelengths. In starburst galaxies the fraction can
be much higher as most of the starlight is absorbed by dust and re-radiated
in the infrared. In extreme sources like the hyperluminous galaxy
$IRAS$\,10214+4724 the energy spectrum peaks around 100$\mu$m in the rest
frame and more than 90\% of the energy is emitted in the infrared and sub-mm
regions. The emission maximum moves further towards sub-mm with increasing
redshift, causing optical studies to seriously underestimate the true star
formation activity. If the dust content is high enough the objects can
remain completely undetected at optical wavelengths.

With ISO and new sub-mm instruments like the SCUBA bolometer array (Holland
et al. \cite{holland99}) it has become possible to study the star formation
history of the universe at infrared to sub-mm wavelengths (for reviews see
e.g. Hughes et al. \cite{hughes3}, \cite{hughes1}). Due to the negative
$K$-correction the observed flux densities will not depend strongly on the
redshift and it is possible to detect more distant galaxies (e.g. van der
Werf \cite{werf}; Guiderdoni et al. \cite{guiderdoni97}).

Recent studies (e.g. Dunlop et al. \cite{dunlop}; Omont et al.
\cite{omont}; Hughes et al. \cite{hughes97}, \cite{hughes3}; Stiavelli et
al. \cite{stiavelli99}; Abraham et al. \cite{abraham99}; Lilly et al.
\cite{lilly99}; Blain et al. \cite{blain99}) have shown that star formation
activity remains high at $z>$1.

In observations with the SCUBA instrument at 450$\mu$m and 850$\mu$m 
(e.g. Hughes et al. \cite{hughes3}; Barger et al. \cite{barger1}; Smail et al.
\cite{smail97}; Blain et al. \cite{blain99}; Eales et al. \cite{eales99};
Lilly et al. \cite{lilly99}) galaxies have been detected up to redshifts
$z$$\sim$5. Compared with galaxies seen in optical surveys the objects have
higher dust content and the star formation rates are an order of magnitude
higher. The surface density of the detected sources exceeds predictions of
no-evolution models by at least one order of magnitude (Smail et al.
\cite{smail97}; Eales et al. \cite{eales99}; Barger et al. \cite{barger99}).
The number of sources detected by Eales et al. (\cite{eales99}) at 850$\mu$m
above $\sim$3\,mJy accounts for $\sim$20\% of the CIRB detected by FIRAS
(Fixsen et al. \cite{fixsen98}). Similar results were obtained by Barger et
al. (\cite{barger99}). At the level of 0.5\,mJy the sources contain most of
the sub-mm CIRB ( Smail et al. \cite{smail97}, \cite{smail98};
Blain et al. \cite{blain99b}).

Kawara et al. (\cite{kawara}) observerved the Lockman Hole at 95$\mu$m
and 175$\mu$m using ISOPHOT. The number of sources found was at least
three times higher than predicted by no-evolution models. The conclusions
of the FIRBACK (Puget et al. \cite{puget99}) and ELAIS (Oliver et al.
\cite{ringberg}) projects are similar and at 175$\mu$m sources with
$S_{\nu}$$>$120\,mJy account for $\sim$10\% of the CIRB detected by
FIRAS.

In this article we study the number density of extragalactic sources and
their contribution to the FIR background radiation using observations made
with ISOPHOT. The data consist of maps made at wavelengths 90$\mu$m,
150$\mu$m and 180$\mu$m, and for some smaller areas at 120$\mu$m. The total
area is close to 1.5 square degrees. Most of the regions have been observed
at three wavelengths (90$\mu$m, 150$\mu$m, 180$\mu$m) some at two
wavelengths (120$\mu$m and 180$\mu$m). Both the galactic foreground cirrus
emission and the emission from typical extragalactic objects will reach
their maxima within or near the observed wavelength range. 
In particular, we will be able to determine the cirrus spectrum for each
region separately.

We have developed a point source extraction routine based on the fitting of
the detector footprint to spatial data. The method is different from those
used in most previous studies where the source detection algorithms have
concentrated on the variations (off-on-off) of the detector signal as
function of time. Our analysis will therefore be independent of and
complementary to the previous results.

\section{Observations and data processing} \label{sect:observations}

\begin{figure*}
\resizebox{\hsize}{!}{\includegraphics{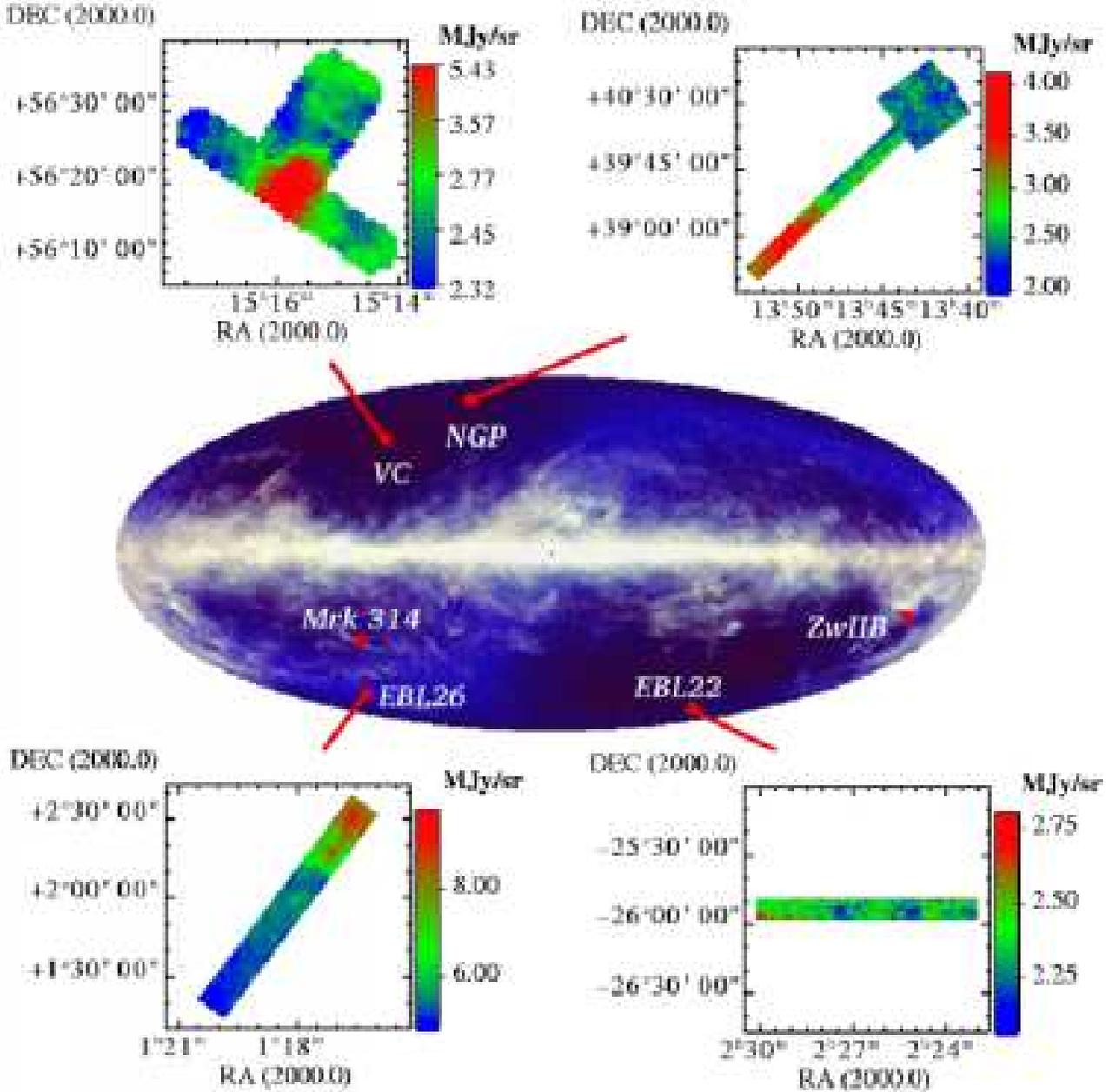}}
\caption[]{
The positions of the observed FIR fields overlaid on the all-sky map composed
of the DIRBE 60, 100, 140 and 240$\mu$m Annual Average Sky Maps. The
180$\mu$m ISOPHOT maps are shown for four regions. Besides dark areas the
fields contain, on purpose, also areas with faint cirrus}
\label{fig:allsky}
\end{figure*}

The observations were performed with the ISOPHOT photometer (Lemke at el.
\cite{lemke}) aboard ISO (Kessler et al. \cite{kessler}). The maps were
made in the PHT22 staring raster map mode using filters C\_90, C\_120,
C\_135, and C\_180 with reference wavelengths at 90$\mu$m, 120$\mu$m,
150$\mu$m and 180$\mu$m, respectively. In this paper we study eight fields
that cover a total of 1.5 square degrees (see Fig.~\ref{fig:allsky}).
Two fields were observed at two wavelengths while the rest have observations
at three wavelengths, the details are shown in Table~\ref{table:maps}. The
C100 detector used for the 90$\mu$m observations consists of 3$\times$3
pixels each with the size of 43.5$\arcsec \times$43.5$\arcsec$ on the sky.
The C200 detector used in the rest of the observations has a raster of
2$\times$2 detector pixels, 89.4$\arcsec \times$89.4$\arcsec$ each.

We have selected regions with low surface brightness. Some maps have
redundancy i.e. the observed pixel rasters partly overlap each other. 
In four maps observed with filter C\_90 the raster step is
larger than the size of the detector, leading to incomplete sampling.

The data were first processed with PIA (PHT Interactive Analysis) program
versions 7.1 and 7.2. Special care was taken to remove glitches caused by
cosmic rays since these might be erroneously classified as point sources
during later analysis. The flux density calibration was made using the FCS
(Fine Calibration Source) measurements (FCS1) performed before and after
each map. Generally the accuracy of the absolute calibration is expected to
be better than 30\% (Klaas et al. \cite{klaas98}). The calibration was
normally applied to observations using linear interpolation between the two
FCS measurements.

The data reduction from the ERD (Edited Raw Data; detector read outs in
Volts) to SCP (Signal per Chopper Plateau; signal at each sky position in
units V\,s$^{-1}$) was performed also using the so-called pairwise method
(Stickel, private comm.). Instead of making linear fits to the ramps
consisting of the detector read-outs one examines the distribution of the
differences between consecutive read-outs. The mode of the distribution is
estimated with myriad technique (Kalluri \& Arce \cite{kalluri98}) and is
used as the final signal for each sky position. This processing was done in
batch mode. Compared with the previous PIA analysis there were some
calibration differences which were possibly due to the different drift
handling of the FCS measurements. For this reason the final surface
brightness values in the new AAP (Astrophysical Applications Data) files
were rescaled using the results of the previous interactive analysis. The
subsequent analysis was carried out with both data sets but no significant
differences were found in the results. The pairwise method is, however,
believed to be more robust against glitches and in the following the results
are based on the data reduced with this method.

\begin{table*}
\caption[]{The eight fields studied here. The columns are:
(1) the name of the field used in this paper, (2)-(3) coordinates of the
centre of each field, (4)-(5) galactic coordinates of the field, (6) area of
the map, (7) the number of raster positions observed over the map, (8) the
step between adjacent raster positions in the staring mode mapping and (9)-(12)
the integration time for each filter. The distance of adjacent scans was in all
cases identical to the raster step used along the scan line}
\label{table:maps}
\begin{tabular}{lrrrrrrrrrrr}
Field  &  \multicolumn{4}{c}{Map Centre}  &  Area &  Rasters  & Step &
          \multicolumn{4}{c}{$t_{\rm int}$ (s)}   \\
	  \cline{2-5} \cline{9-12}
       &    RA(2000.0) & DEC(2000.0)   & $l$ & $b$ &  (\sq \degr) &      &  
       ($\arcsec)$  &       C\_90 & C\_120 & C\_135 & C\_180  \\
(1) & (2) & (3) & (4) & (5) & (6) & (7) & (8) & (9) & (10) & (11) & (12) \\      
\hline
VCN    & 15 15 21.7 &  +56 28 58  & 91.76 & 51.42 &  0.030 &  10$\times$4    &
90  & 46  &    & 46  & 46  \\	
VCS    & 15 15 53.1 & +56 19 30   & 91.27 & 51.40 &  0.023 &  21$\times$2    &  
90  & 46  &    & 46  & 46  \\	
NGPN   & 13 43 53.0 & +40 11 35   & 86.82 & 73.61 & 0.27   &  32$\times$4    &  
180 & 23  &    & 27  & 27  \\
       & 13 42 32.0 & +40 29 06   & 87.88 & 73.26 & 0.53   &  15$\times$15   &  
       180 &     &    &     & 32  \\
NGPS   & 13 49 43.7 & +39 07 30   & 81.49 & 73.30 & 0.27   &  32$\times$4    &  
180 & 23  &    & 27  & 27  \\
EBL22  & 02 26 34.5 & -25 53 43   & 215.78 & -69.19 & 0.19 &  32$\times$3    &  
180 & 23  &    & 27  & 27  \\
EBL26  & 01 18 14.5 & 01 56 40    & 135.89 & -60.66 & 0.27 &  32$\times$4    &  
180 & 23  &    & 23  & 23  \\
Mrk\,314 & 23 02 58.7 & 16 38 09    & 88.29 & -38.90 & 0.045 & 6$\times$6    &  
135 &     & 36 &    &  36  \\
ZW\,IIB     & 05 10 46.6 & -02 43 21   & 203.54 & -24.02 & 0.018 & 5$\times$3 & 
135 &     & 36 &    &  36  \\
\end{tabular}
\end{table*}

\section{The point source detection} \label{sect:method}

The point source detection was performed in two steps using data processed
to the AAP level with PIA and the pairwise method (see
Sect.~\ref{sect:observations}). The data consists of surface brightness
values with error estimates and were flat fielded using special routines
(see Appendix \ref{appendix:flatfield}).

In the first phase each surface brightness value was compared with the mean
and the standard deviation estimated from other measurements within a fixed
radius. The radius was set to about three times the size of the detector
pixel. Values rising above the local mean surface brightness by more than
0.7$\sigma$ were considered as potential point sources.

In the second phase a model consisting of point source and a constant
background was fitted into each region surrounding the candidate positions.
In the fit the footprint matrices were used to calculate the contribution of
the point source to the observed surface brightness values. The free
parameters of the fit were the source flux density, the two coordinates of
the source position, and the background surface brightness. Details of the
fitting procedure are given in Appendix \ref{appendix:fitting}

\subsection{Completeness and false detections} \label{sect:false}

As a result of the fitting procedure we obtained estimates for the flux
density, coordinates of the potential point source and a value for the local
background brightness. The fitting routine provides formal error estimates
for all free parameters. In addition, we have calculated the standard
deviation of the surface brightness values inside the selected region.
Assuming that the flux density values are normally distributed we can
calculate for each source candidate the probability, $P$, that the detection
is not caused by background noise.

The completeness of the point source detection and the probability of false
detections were studied with simulations (see Appendix
\ref{appendix:false}). The number of false sources was found to somewhat
exceed the expected number of 1-$P$ per map pixel and this was taken into
account in setting the selection criteria. A confidence limit of $P$=99\%
(corresponding to $\sim$2.3\,$\sigma$) was used to discard uncertain
detections.

Simulated measurements were also used to study the effects of imperfect flat
fielding (Appendix \ref{appendix:fferror}). These do not cause significant
errors in the source counts. Cosmic ray glitches and short term detector
drifts may still lead to spurious detections primarily on the C100 detector.
Glitches have been removed during the standard data reduction and any
remaining anomalies should be reflected in the error estimates calculated
for the surface brightness values. Large error estimates will reduce the
number of false detections (see
\ref{appendix:visual}).

For the purpose of source counts (see Sect.~\ref{sect:count}) we will use an
{\em additional} criterion based on the ratio between the source flux
density, $S_{\nu}$, given in Jy and the background rms noise, $\sigma_{\rm
bg}$, given in units of Jy per pixel,
\begin{equation}
\rho = \frac{S_{\nu}}{\sigma_{\rm bg}}.
\end{equation}
The parameter $\rho$ is not directly related to the probability obtained
from the footprint fit and can be used as safeguard against
false detections. In Sect.~\ref{sect:count} we will use a limit
$\rho>\rho_{\rm 0}=$10.5. In the maps fluctuations are typically below
0.1\,MJy\,sr$^{-1}$ which, in the case of the C200 detector, corresponds to
$\sigma_{\rm bg}\sim$0.02\,Jy per pixel. Our $\rho$-criterion implies thus a
typical detection limit of 200\,mJy. Because of the smaller pixel size of
the C100 detector, $\sim$44$\arcsec$ instead of $\sim$89$\arcsec$, the
detection limit is lower by a factor of four. Note that since the source
flux is always distributed over several map pixels $\rho$ cannot be
interpreted directly as a $\sigma$ limit.

The value $\rho_{\rm 0}=$10.5 was chosen based on simulations: at this limit
the detection rate is $\sim$70\% for sources which fulfill the previous
criterion of $P>$99\%. The number of false detections is less than one per
200 map pixels. For the C200 detector the expected number of false
detections within the mapped areas is $\la$10 i.e. {\em less} than the
probable number of true sources above the $\rho$ limit that remain
undetected. Because of the larger number of map pixels in the 90$\mu$m maps
there can be {\em more} false detections than undetected true sources.
However, the relative error in the source counts should remain well below
50\%.

\subsection{Multiwavelength confirmation of source detections}
\label{sect:multi}

Since the source counts might be slightly overestimated even after applying
the additional $\rho$-criterion constraint we have constructed another
source list based on multi-wavelength detections. A source is accepted only
if there are detections at two wavelengths, each at 99\% confidence level
(see above), and the spatial distance between these detections
is $\le$80$\arcsec$. The $\rho$ criterion is not used. If a map contains
several possible sources within that distance the one with the highest
estimated probability is selected. With the number density of detections at
the 99\% confidence level we can estimate the probability of finding by
accident a source within 80$\arcsec$ radius of any given position: 5\% for
the 90$\mu$m and 150$\mu$m maps and below 10\% even for 180$\mu$m.

We can calculate a rough estimate for the number of false detections as
follows. For any individual detection we have an 1\% probability that it is
not caused by a real source. Empirically, the probability of finding another
detection at a different wavelength within 80$\arcsec$ radius is less than
10\%. This increases the confidence in the first detection from 99\% to
99.9\%. In the source counts for 150$\mu$m and 180$\mu$m the total number of
spurious detections can therefore be estimated to be $\sim$1. Because of the
larger number of pixels in the 90$\mu$m maps the number of false 90$\mu$m
sources could be up to $\sim$6.
However, the false detections are made only close to the background noise
level where the number of undetected true sources can easily exceed this.
Counting only detections with multiwavelength confirmation will
underestimate the true number of sources. The probability of finding two
spurious detections within 80$\arcsec$ is 0.001\% and the confidence level
of a source detected at two wavelengths (instead of a detection at one
wavelength only) equals a 4.2$\sigma$ detection.

Examples of such multiwavelength detections are shown in
Fig.~\ref{fig:ebl26map}. The positions of four sources in the field EBL26
are overlaid on the 90$\mu$m, 150$\mu$m and 180$\mu$m maps. Sources 5 and 6
are detected only at 150 and 180\,$\mu$m (see also
Table~\ref{table:sources}). The leftmost source is a 90$\mu$m detection only
and is therefore not in Table~\ref{table:sources}. It is probably caused by
a glitch as can be seen from the SRD data (Signal per Ramp Data i.e.
signals, V\,s$^{-1}$, derived from individual detector integration ramps) in
Fig.~\ref{fig:ebl26srd}a. The signals are shown for four pixels at five
positions (given as a function of time) centered on the position closest to
the fitted source position. Source 7 was detected only at 150$\mu$m and
180$\mu$m while at 90$\mu$m it remained below our detection limit. The SRD
data for this source are also shown in Fig.~\ref{fig:ebl26srd}. One must
remember that the detection procedure was not based on SRD data and in
Fig.~\ref{fig:ebl26srd} only part of the relevant data are shown. SRD
signals were, however, used to visually estimate the possible effect of
glitches on the source detections (see
\ref{appendix:visual}).

\begin{figure}
\epsfxsize=7.0cm  \epsfbox[60 210 460 560]{1850.f2a}
\epsfxsize=7.0cm  \epsfbox[60 210 460 560]{1850.f2b}
\epsfxsize=7.0cm  \epsfbox[60 210 460 560]{1850.f2c}
\caption[]{
The positions of four detections in EBL26 overlaid on the 90$\mu$m,
150$\mu$m and 180$\mu$m maps. The leftmost source was detected only at
90$\mu$m while the other sources (numbers 5, 6 and 7) were detected at
150$\mu$m and 180$\mu$m. See Table~\ref{table:sources} for parameters of the
numbered sources}
\label{fig:ebl26map}
\end{figure}

\begin{figure*}
\resizebox{\hsize}{!}{\includegraphics{1850.f3}}
\caption[]{
SRD signals for two sources shown in Fig.~\ref{fig:ebl26map}: a source
detected at 90$\mu$m only ({\bf a}) and the Source number 7 ({\bf b}). The
signals are shown for four pixels as function of time for five consecutive
raster positions. For each pixel the plot is centered on the position
closest to the average positions of the detections made at different
wavelengths. The signals from this position are between the dashed lines.
The pixel numbers are shown in the figure. The first source was detected
at 90 $\mu$m only while the second one was detected at 150$\mu$m and at
180$\mu$m}
\label{fig:ebl26srd}
\end{figure*}

For most of the area of the 15$\times$15 position 180$\mu$m map NGPN no
observations at other wavelengths were available. In order to confirm the
detections the data were divided at the ERD level into two parts. The first
data set contained the first ramps of each measurement and the second data
set the rest of the ramps. These data sets were used to create two maps and
sources were detected and accepted according to procedures described above.
The maps are not completely independent. A large glitch that affects several
ramps may have influenced both data sets. The risk for false detections is
therefore larger than in other cases. Some of the detections could, however,
be confirmed with observations of the 32$\times$4 position NGPN map which
partly overlaps the square 15$\times$15 position 180$\mu$m map (see
Table~\ref{table:maps}).

\section{Discrimination against cirrus knots}
\label{sect:spectra}

We will check the possibility that some of the sources detected are
small scale cirrus structures (cirrus knots).

\subsection{Cirrus confusion noise} \label{sect:confusion}

We estimate the contribution of cirrus to the sky confusion. The total
noise, $\sigma_{\rm total}$, in the source flux densities is estimated from
the standard deviation of the flux densities. This is obtained by fitting a
point source to the position of each measurement (each pixel and raster
position) while keeping the source positions fixed i.e. effectively
convolving the map with the detector footprint. The total noise consists of
measurement noise, $\sigma_{\rm meas.}$, from the detector, and the sky
confusion, $\sigma_{\rm sky}$, caused by real sky brightness variations,
\begin{equation}
\sigma_{\rm total}^2 = \sigma_{\rm meas.}^2 + \sigma_{\rm sky}^2.
\label{eq:noise}
\end{equation}
The measurement noise, in flux density units, is derived from the error
estimates obtained for the surface brightness values from the SRD data. The
surface brightness map was modified by adding corresponding amount of
normally distributed noise. The fitting procedure was then repeated. The rms
difference between the flux densities obtained from the two maps gives an
estimate of measurement errors on the flux density scale.
Table~\ref{table:noises} summarizes the results for the fields EBL26 and
NGPS. The sky confusion is obtained (using Eq.\ref{eq:noise}) from the
total and measurement noise estimates. The measurement noise was seen to be
roughly equal to the sky confusion. In order to estimate the contribution of
cirrus to the sky noise we have used the approach of Gautier et al
(\cite{gautier92}).

\begin{table}
\caption[]{The estimated total rms noise, $\sigma_{\rm total}$,
in the fields EBL26 and NGPS compared with the estimated measurement noise,
$\sigma_{\rm meas.}$, and the sky confusion, $\sigma_{\rm sky}$  }
\label{table:noises}
\begin{tabular}{lrrrrrrrrr}
field  &  $\lambda$  & $\sigma_{\rm total}$  & $\sigma_{meas.}$ &
$\sigma_{\rm sky}$  \\
     &  ($\mu$m)   & (mJy)        & (mJy)             & (mJy)          \\
\hline
EBL26 &  90        &   41         &  32               &  25            \\
      & 150        &  117         &  55               & 104            \\
      & 180        &  110         &  56               &  88            \\
NGPS  &  90        &   29         &  24               &  17            \\
      & 150        &   51         &  36               &  36            \\
      & 180        &   57         &  41               &  40            \\
\end{tabular}
\end{table}

Gautier et al. (\cite{gautier92}) calculated the confusion noise due to
infrared cirrus for different observation strategies. We use these results
to estimate cirrus contamination in the case of circular aperture with
diameter $d$, immediately surrounded by a reference annulus of the same
width. This configuration is not exactly the same as in our detection
procedure but should give accurate estimates of the expected cirrus
confusion.

Herbstmeier et al. (\cite{herbstmeier}) estimated for the large NGPN
180$\mu$m map a fluctuation power $P=$2.3$\times$10$^3$\,Jy$^2$\,sr$^{-1}$ at the
scale of 4$\arcmin$. The fluctuations are mostly due to cirrus emission.
With the dependence $P\sim d^2$ derived by Herbstmeier et al. we obtain a
fluctuation amplitude of 18\,Jy\,sr$^{-0.5}$ at $d=$1.5$\arcmin$ which
corresponds to the size of the detector beam. According to the tables of
Gautier et al. (\cite{gautier92}) the estimated flux density fluctuations
due to cirrus emission are 8.0\,mJy. In this field all detected sources have
flux densities exceeding 100\,mJy. It is therefore very unlikely that
they could be caused by cirrus.

The surface brightness attributed to cirrus is obtained by subtracting the
CIRB given by Fixsen et al. (\cite{fixsen98}), 0.82\,MJy\,sr$^{-1}$ at
180$\mu$m, and the zodiacal light according to Leinert et al.
(\cite{leinert98}). In the field, EBL26, which has the largest surface
brightness among our fields, the average cirrus surface brightness is
$\sim$1.4\,MJy\,sr$^{-1}$ compared to $\sim$0.8\,MJy\,sr$^{-1}$ in the NGPN field. With
the relation $P\sim<B>^3$ (Gautier et al. \cite{gautier92}) between the
fluctuation power and the mean surface brightness we can estimate that in
EBL26 the cirrus fluctuation level is $\sim$43\,mJy. All sources are above
$\sim$150\,mJy, i.e. above the 3.5\,$\sigma$ level.

The results of Herbstmeier et al. (\cite{herbstmeier}) show that at the
scale of the C100 beam size, $d\sim45\arcsec$, the expected cirrus
fluctuation amplitude is clearly below 10\,Jy\,sr$^{-0.5}$ for all our
90$\mu$m maps. According to Gautier et al. (\cite{gautier92}) this
corresponds to a flux density of 4\,mJy which is again clearly below the
flux densities of even the faintest sources at 90\,$\mu$m in the present
study.

In conclusion, it is clear that cirrus is not the dominant factor in the sky
confusion noise. The flux density values derived for cirrus contamination
are small compared with the faintest sources detected here. Because of the
non-gaussian nature of the cirrus fluctuations (Gautier et al.
\cite{gautier92}) it is, however, possible that our source lists contain
some cirrus knots.

We have, however, also the advantage of having observations at different
wavelengths which offers an additional means for the discrimination against
spurious detections caused by galactic cirrus.

\subsection{Cirrus spectra} \label{sect:cirrus}

Using observations made at different wavelengths we can determine the cirrus
spectrum in each region. For each 180$\mu$m measurement the corresponding
surface brightness values at the other wavelengths were calculated using
weighting with a gaussian with the approximate size of the C200 detector
beam. Linear fits were performed to surface brightness values at
90$\mu$m or 150$\mu$m plotted against the 180$\mu$m values and the slopes
were used to derive the cirrus spectrum (see Juvela et al. \cite{vilspa}).

The spectra obtained are shown in
Figs.~\ref{fig:90det_vc}-\ref{fig:90det_ngp}. The cirrus spectrum can be
determined most reliably in regions with clear surface brightness variations
which is the case for fields EBL26 and NGPS. It is also, nevertheless,
possible to determine the spectra for all other fields. Bright individual
sources were always removed from the data but in the VCS and VCN fields the
results may be influenced by the emission from the nearby galaxy, NGC\,5907.

\subsection{Source spectra vs. cirrus spectra} \label{sect:knots}

The cirrus spectra were compared with the spectra of the detected sources.
The relationship between the cirrus surface brightness spectrum and the
``source'' spectrum due to a cirrus knot depends on the size of the cirrus
knot: (1) If the cirrus knot is small compared with the C100 detector pixel,
the knot spectrum is the same as the surface brightness spectrum. (2) If the
knot size is $>$45$\arcsec$ some part of the 90$\mu$m flux is included into
the background, and the source flux density at 90$\mu$m drops. (3) If the
cirrus knot is large compared even with the pixel size of the C200 detector
also at the longer wavelengths only part of the total flux density contained
in the cirrus knot will be detected. In that case there will be probably no
detection with the C100 detector.

Simulations showed that for gaussian cirrus knots with FWHM
$\sim$90$\arcsec$ we detect with the C100 detector less than half of the
total flux density. For cirrus knots with FWHM$\sim$180$\arcsec$ the ratios
$F(90\mu$m)/$F(150\mu$m) and $F(90\mu$m)/$F(180\mu$m) drop to $\la$0.25 of
the cirrus surface brightness values.

If the source spectrum is flat compared to the cirrus surface brightness
spectrum we can conclude the source is not a cirrus knot. Otherwise, this
alternative cannot be excluded with certainty. However, when the flux
density ratio $F(90\mu$m)/$F(150\mu$m) or $F(90\mu$m)/$F(180\mu$m) is
clearly larger than one fourth of the corresponding ratio in the cirrus
surface brightness spectrum it is improbable that the source is due to a
cirrus knot because this would require that the knot is much smaller than
180$\arcsec$. From the slope of the cirrus power spectrum as function of
spatial frequency, (Gautier et al.~\cite{gautier92}; Low \& Cutri~\cite{low};
Herbstmeier et al. \cite{herbstmeier}) and the very low level of cirrus
emission in most of the fields, we estimate that it is very improbable that
there are many small ($\la$90$\arcsec$) cirrus knots strong enough to be
detected as sources (see Sect.~\ref{sect:confusion}).

\begin{figure}
\resizebox{\hsize}{!}{\includegraphics{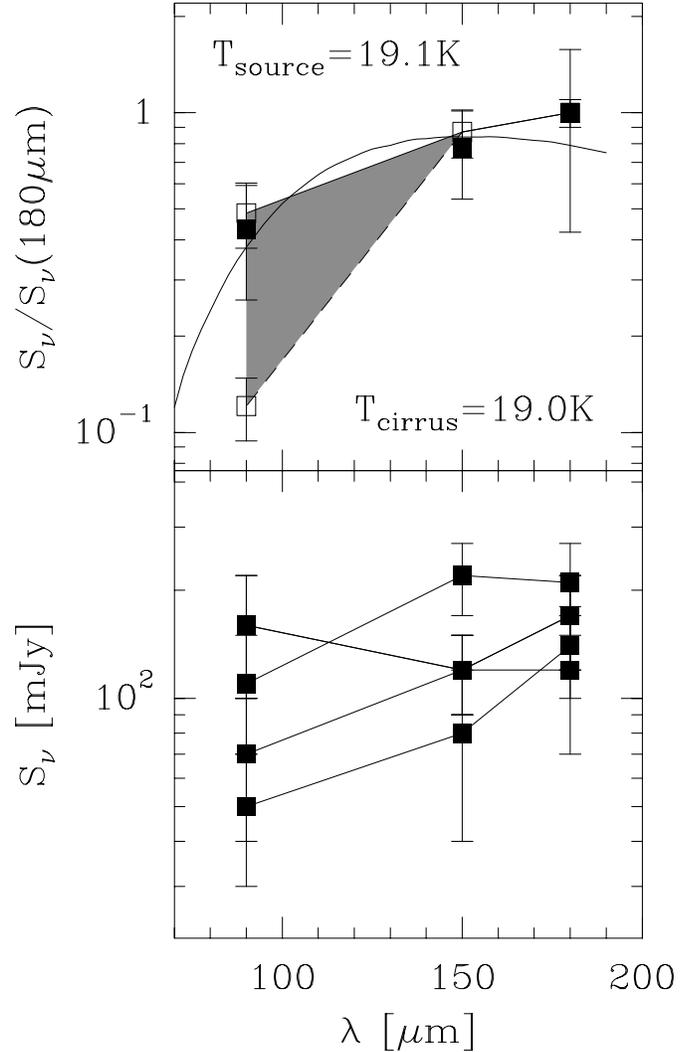}}
\caption[]{
Average source spectrum of the fields VCN and VCS (filled squares) and the
cirrus spectrum determined in the same fields (open squares). The shaded
area indicates the possible range for cirrus spectra. At 90$\mu$m the
highest cirrus flux is obtained when the knot is small compared with the
C100 pixel size and the lowest flux density corresponds to a knot with size
similar to the C200 detector pixels. The upper cirrus spectrum and the
average source spectrum have been fitted with modified Planck Law, $\nu^2
B_{\nu}$, with temperatures shown in the figure {\em (upper frame)}. Spectra
of sources in fields VCN and VCS detected at 90$\mu$m and at one or two
longer wavelengths. The average source spectrum has been calculated without
including the source with the highest 90$\mu$m flux density ({\em lower
frame}) }
\label{fig:90det_vc}
\end{figure}

\begin{figure}
\resizebox{\hsize}{!}{\includegraphics{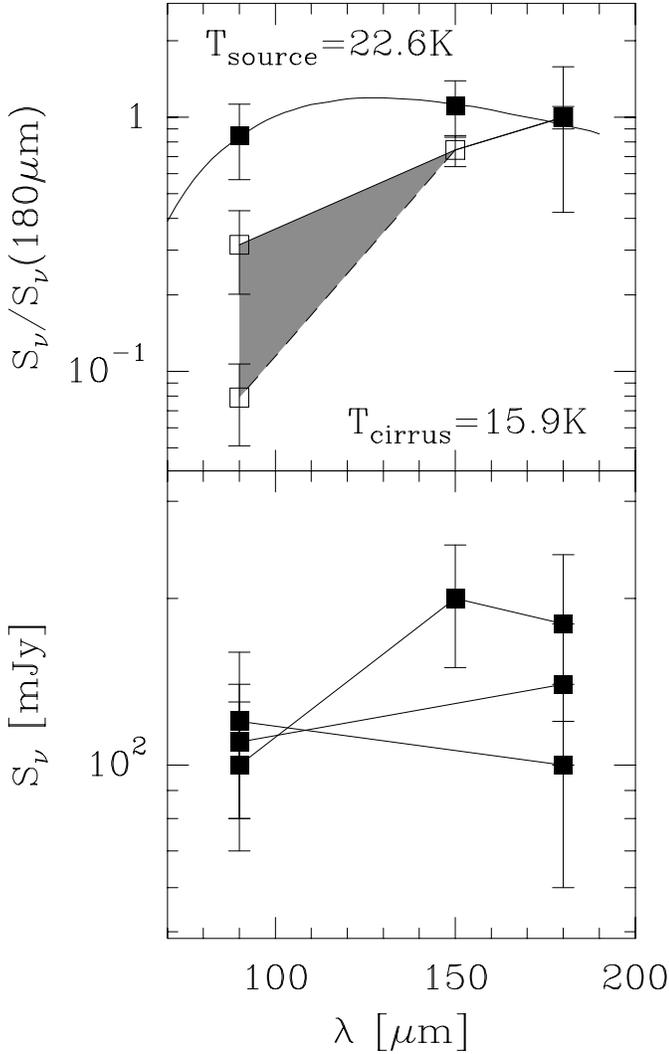}}
\caption[]{Cirrus spectrum and source spectra in the field EBL22 (symbols
as in Fig.~\ref{fig:90det_vc}}
\label{fig:90det_ebl22}
\end{figure}

\begin{figure}
\resizebox{\hsize}{!}{\includegraphics{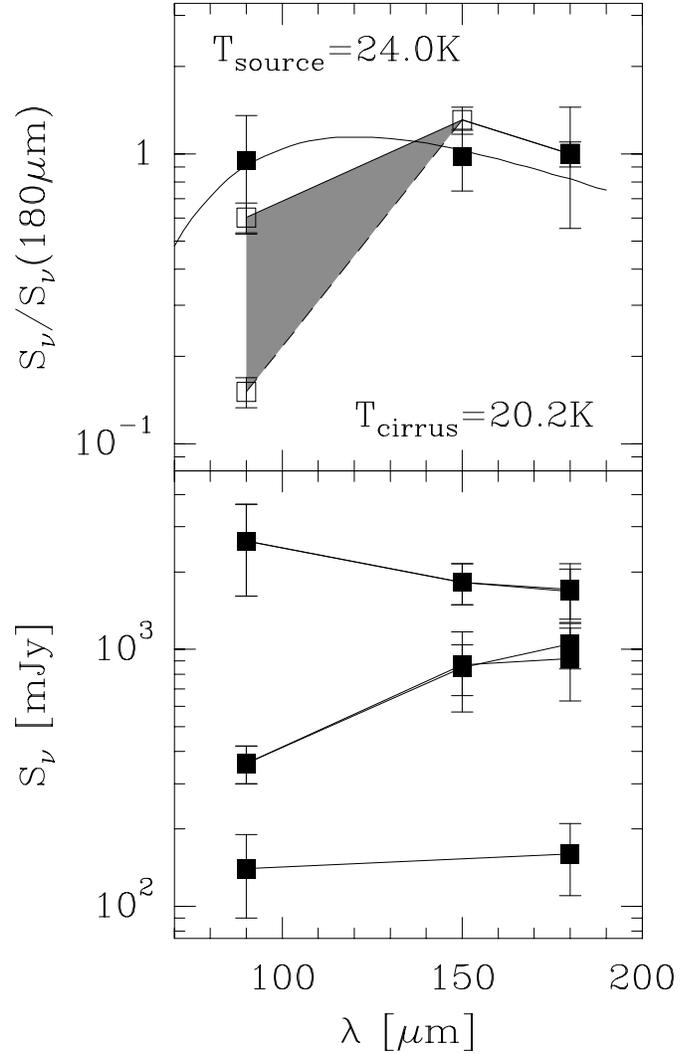}}
\caption[]{Cirrus spectrum and source spectra in the field EBL26 (symbols
as in Fig.~\ref{fig:90det_vc}}
\label{fig:90det_ebl26}
\end{figure}

\begin{figure}
\resizebox{\hsize}{!}{\includegraphics{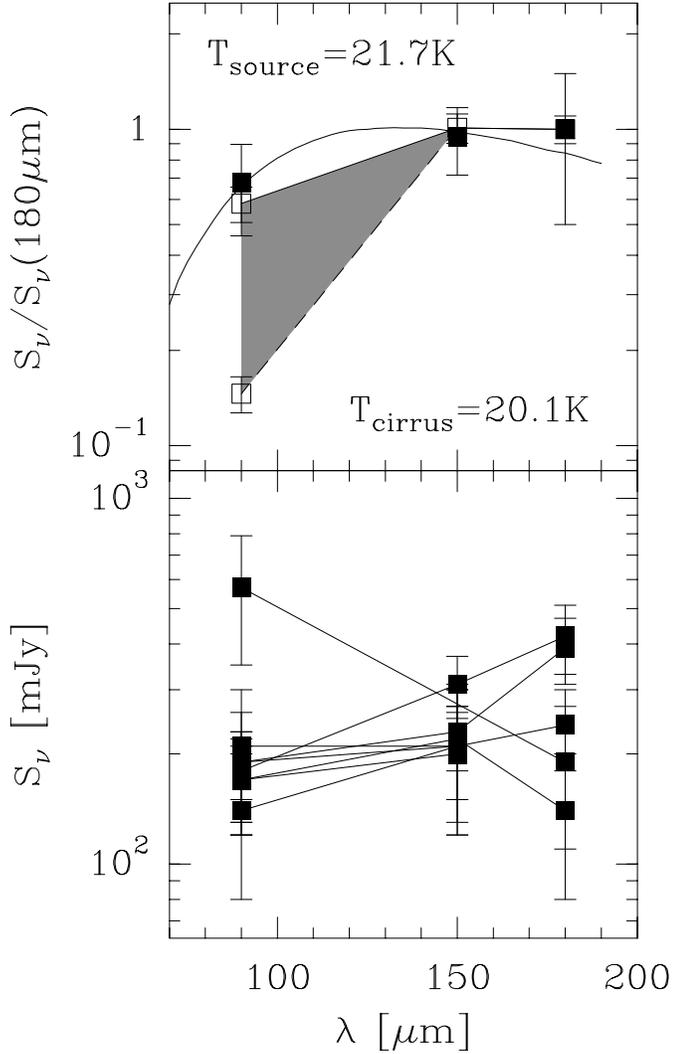}}
\caption[]{Cirrus spectrum and source spectra combined from the fields NGPN
and NGPS (symbols as in Fig.~\ref{fig:90det_vc}. The plotted average
source spectrum has been calculated without including the source with the
highest 90$\mu$m flux}
\label{fig:90det_ngp}
\end{figure}

In Figs.~\ref{fig:90det_vc}-\ref{fig:90det_ngp} we show for our four main
fields the predicted spectra caused by cirrus knots together with those
actual sources that were detected at 90$\mu$m and at some longer wavelength.
The cirrus spectra have been drawn for two cases assuming either that the
cirrus knot is smaller than the C100 detector pixel (solid line) or that it
is of the same size as the beam of the C200 detector pixels (dashed line).

The source spectra are almost without exception flat compared with the
steeper of the two cirrus spectra and the sources are not likely to be
caused by cirrus. Furthermore, in many cases the relative 90$\mu$m flux is
higher than what is possible even for very small cirrus knots. No sources
were rejected from our sample based on their SED shape.

\begin{figure}
\resizebox{8cm}{!}{\includegraphics{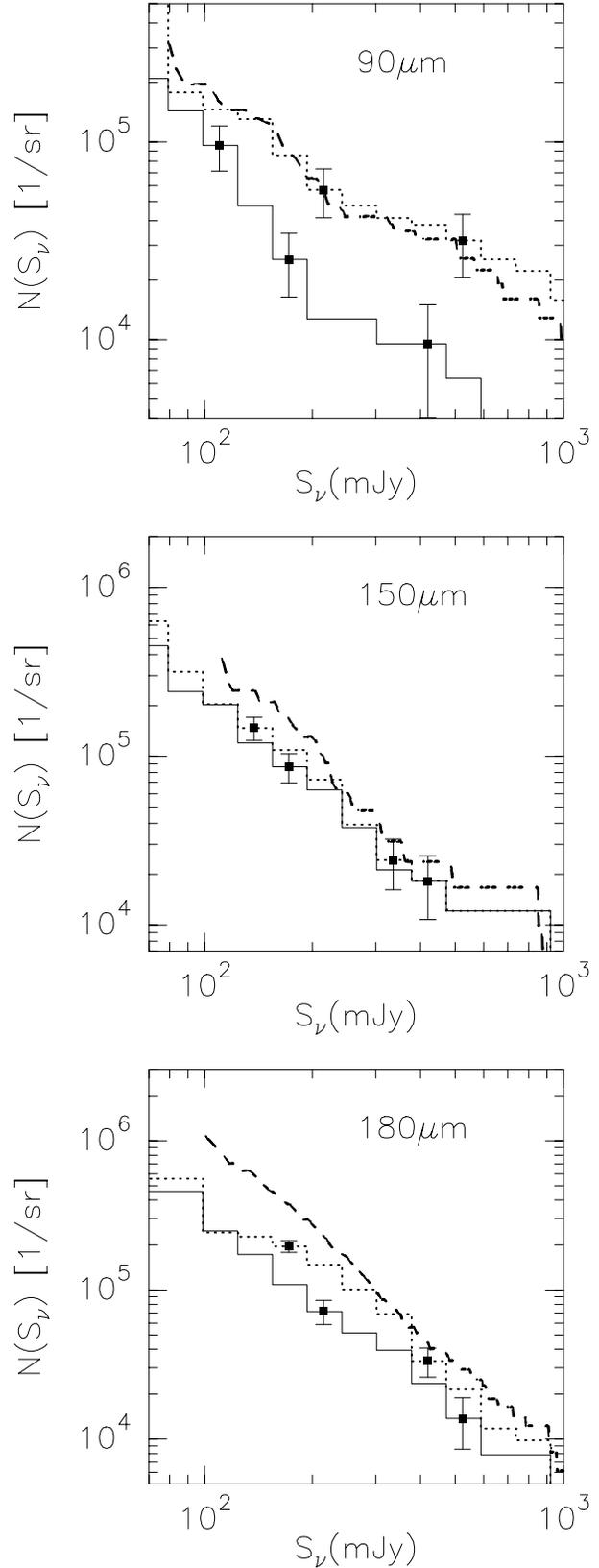}}
\caption[]{
Cumulative source counts at 90$\mu$m, 150 $\mu$m and 180$\mu$m. The solid
and the dotted lines represent values obtained by estimating the areas
according to the faintest sources detected in a map. The dotted line gives
all sources detected and the solid line sources detected at more than one
wavelength. The dashed curves represent cases in which both the the source
selection and the area determination were based on the local background
variations (see Sect.~\ref{sect:count}). The Poisson noise of the count
statistics is indicated at some representative points with vertical
error bars

}
\label{fig:counts}
\end{figure}

\section{Source counts and identifications}

\subsection{Cumulative source counts} \label{sect:count}

Table~\ref{table:sources} contains sources detected at two or three
wavelengths with an individual detection probability $P>$99\% at one
wavelength (see Sect.~\ref{sect:multi}). The coordinates are averages,
weighted with the estimated confidence of detections at different
wavelengths. The flux densities and their error estimates are obtained from
the fitting routine. If the field was mapped at three wavelengths but the
source was detected only at two wavelengths, then we quote an upper limit
based on the background surface brightness variations, $S=10.5\sigma_{\rm
bg}$ where $\sigma_{\rm bg}$ is given in units of Jy per pixel. The
detection probability corresponding to this limit is 97\%.

\begin{table*}
\caption[]{(See next page)
Sources detected at two or more wavelengths. The coordinates are weighted
averages of the positions fitted at different wavelengths. For sources
undetected at one wavelength we give upper limits which, according to
simulations, correspond to the flux density of a source with detection
probability 97\%. Quality flags, $q$, are based on visual inspection of the
SRD data and the scale is from 0 (obvious glitch) to 4 (clear source).
Counts $c$ are the results from our detection procedure based on the SRD
data (see \ref{appendix:visual}). Asterisks mark cases where a detection at
one wavelength could be associated with two different detections at another
wavelenth. The confidence level of each detection at one wavelength is
99\% while the confidence level for a source with two detections at different
wavelengths is better than 99.99\%
}
\end{table*}

\setcounter{table}{3}

\begin{table*}
\label{table:sources}
\begin{tabular}{lccccccccccc}
&  &    &    & \multicolumn{2}{c}{90$\mu$m} & \multicolumn{2}{c}{120$\mu$m} 
  & \multicolumn{2}{c}{150$\mu$m} & \multicolumn{2}{c}{180$\mu$m} \\
Field & Source No. & R.A.  &  Dec  &  $S$  & $q/c$ &  $S$ & $q/c$ &  $S$ &  $q/c$ &  $S$ &  $q/c$ \\
(2000.0) & (2000.0) & (Jy) & & (Jy) & & (Jy) & & (Jy) & \\
\hline
EBL22
 & 1 &  02 24 27.2 &  -25 55 11 &  $<$0.10   &     &     -      &      &   0.12(0.04) & 3/7 &   0.15(0.05) & 4/10\\
 & 2 &  02 24 33.6 &  -25 50 18 &  $<$0.07   &     &     -      &      &   0.23(0.03) & 4/10 &   0.21(0.05) & 4/10\\
 & 3 &  02 25 28.0 &  -25 56 34 &  $<$0.07   &     &     -      &      &   0.21(0.05) & 3/10 &   0.27(0.05) & 4/10\\
 & 4 &  02 26 58.4 &  -25 50 49 &   0.12(0.04) & 3/19 &  -      &      &  $<$0.13   &     &   0.10(0.04) & 2/2\\
 & 5 &  02 28 19.7 &  -25 56 55 &   0.10(0.03) & 2/16 &  -      &      &   0.20(0.05) & 3/4 &   0.18(0.06) & 2/6\\
 & 6 &  02 28 44.8 &  -25 50 31 &   0.11(0.03) & 1/15 &  -      &      &  $<$0.14   &     &   0.14(0.04) & 3/8\\
EBL26
 & 1$^*$ &  01 16 38.9 &  +02 30 45 &   0.36(0.06) & 1/4 &    -    &      &   0.85(0.19) & 4/9 &   1.05(0.21) & 4/7\\
 & 2$^*$ &  01 16 40.3 &  +02 30 01 &   0.36(0.06) & 1/4 &    -    &      &   0.87(0.30) & 4/8 &   0.92(0.29) & 4/7\\
 & 3$^*$ &  01 17 09.2 &  +02 16 45 &   2.64(1.03) & 4/20 &   -    &      &   1.82(0.33) & 4/10 &   1.68(0.37) & 4/8\\
 & 4$^*$ &  01 17 09.3 &  +02 17 13 &   2.64(1.03) & 4/20 &   -    &      &   1.82(0.33) & 4/10 &   1.71(0.44) & 4/8\\
 & 5 &  01 17 49.7 &  +02 00 29 &  $<$0.14   &     &    -       &      &   0.33(0.13) & 4/10 &   0.36(0.10) & 3/9\\
 & 6 &  01 17 53.5 &  +02 09 25 &  $<$0.22   &     &    -       &      &   0.48(0.14) & 4/10 &   0.34(0.10) & 3/9\\
 & 7 &  01 18 03.8 &  +01 59 09 &  $<$0.13   &     &    -       &      &   0.85(0.14) & 4/10 &   0.56(0.08) & 4/10\\
 & 8 &  01 18 34.6 &  +01 45 07 &  $<$0.16   &     &    -       &      &   1.04(0.10) & 4/10 &   0.59(0.11) & 3/10\\
 & 9 &  01 19 07.5 &  +01 31 15 &   0.14(0.05) & 2/14 &  -      &      &  $<$0.26   &     &   0.16(0.05) & 2/5\\
Mrk\,314
 & 1 &  23 02 52.3 &  +16 31 52 &       -     &      &   0.20(0.06) & 1/4 &      -      &      &   0.32(0.11) & 3/8\\
 & 2$^1$ &  23 02 59.7 &  +16 36 10 &   -         &      &   1.32(0.17) & 2/7 &  -          &      &   0.91(0.17) & 4/10\\
 & 3 &  23 03 30.6 &  +16 36 11 &       -     &      &   1.09(0.16) & 2/2 &      -      &      &   1.09(0.19) & 2/4\\
NGP 
 & 1 &  13 41 22.8 &  +40 41 54 &  $<$0.17   &     &      -     &      &   0.49(0.06) & 2/6 &   0.50(0.06) & 3/6\\
 & 2 &  13 42 03.1 &  +40 28 13 &  $<$0.09   &     &      -     &      &   0.31(0.06) & 3/7 &   0.27(0.05) & 3/6\\
 & 3 &  13 42 22.1 &  +40 21 58 &   0.17(0.05) & 3/11 &   -     &      &   0.20(0.07) & 2/0 &  $<$0.30   &    \\
 & 4 &  13 42 41.7 &  +40 27 12 &  $<$0.13   &     &      -     &      &   0.23(0.05) & 3/5 &   0.29(0.07) & 2/6\\
 & 5 &  13 42 46.9 &  +40 18 02 &  $<$0.14   &     &      -     &      &   0.30(0.08) & 2/8 &   0.45(0.10) & 3/8\\
 & 6 &  13 43 03.3 &  +40 14 51 &   0.19(0.07) & 3/15 &   -     &      &   0.23(0.08) & 2/0 &   0.39(0.08) & 2/1\\
 & 7$^*$ &  13 43 39.8 &  +40 13 56 &   0.14(0.06) & 3/14 & -      &      &   0.21(0.06) & 3/5 &   0.24(0.06) & 3/10\\
 & 8$^*$ &  13 43 41.5 &  +40 14 02 &  $<$0.13   &     &    -      &      &   0.21(0.06) & 3/5 &   0.24(0.06) & 3/10\\
 & 9 &  13 43 46.7 &  +40 07 54 &  $<$0.11   &     &        -   &      &   0.14(0.06) & 1/0 &   0.20(0.07) & 2/0\\
 & 10 &  13 45 03.2 &  +39 55 04 &  $<$0.13   &     &       -   &      &   0.18(0.06) & 4/10 &   0.25(0.10) & 4/10\\
 & 11 &  13 45 44.6 &  +39 47 17 &  $<$0.10   &     &       -   &      &   0.13(0.05) & 4/9 &   0.21(0.08) & 4/10\\
 & 12$^*$ &  13 47 17.3 &  +39 36 50 &   0.19(0.04) & 3/13 &  -    &      &   0.21(0.09) & 2/3 &  $<$0.23   &    \\
 & 13$^*$ &  13 47 19.0 &  +39 37 00 &   0.21(0.09) & 3/10 &  -    &      &   0.21(0.09) & 2/3 &  $<$0.26   &    \\
 & 14 &  13 47 33.8 &  +39 32 04 &  $<$0.15   &     &       -   &      &   0.19(0.05) & 3/1 &   0.22(0.09) & 2/0\\
 & 15 &  13 48 24.0 &  +39 21 23 &  $<$0.13   &     &       -   &      &   0.16(0.05) & 2/1 &   0.16(0.04) & 2/0\\
 & 16$^*$ &  13 48 30.4 &  +39 27 12 &  $<$0.11   &     &   -      &      &   0.22(0.04) & 3/4 &   0.14(0.06) & 3/8\\
 & 17$^*$ &  13 48 31.9 &  +39 26 58 &   0.17(0.04) & 1/4 & -      &      &   0.22(0.04) & 3/4 &   0.14(0.06) & 3/8\\
 & 18 &  13 49 07.6 &  +39 16 54 &   0.18(0.05) & 2/12 &    -   &      &   0.31(0.06) & 3/6 &   0.42(0.09) & 3/2\\
 & 19 &  13 49 31.5 &  +39 04 55 &   0.57(0.22) & 3/16 &    -   &      &  $<$0.20   &     &   0.19(0.08) & 2/0\\
 & 20 &  13 49 36.1 &  +39 07 11 &  $<$0.15   &     &       -   &      &   0.15(0.06) & 3/6 &   0.20(0.08) & 2/3\\
 & 21 &  13 50 35.8 &  +39 00 29 &  $<$0.14   &     &       -   &      &   0.21(0.08) & 2/1 &   0.26(0.09) & 2/3\\
 & 22 &  13 50 56.2 &  +38 58 18 &  $<$0.12   &     &       -   &      &   0.32(0.11) & 2/2 &   0.42(0.11) & 3/4\\
 & 23 &  13 52 14.1 &  +38 39 35 &  $<$0.11   &     &       -   &      &   0.17(0.06) & 3/6 &   0.21(0.07) & 2/0\\
 & 24 &  13 52 33.1 &  +38 42 33 &  $<$0.13   &     &       -   &      &   0.26(0.06) & 4/8 &   0.28(0.08) & 3/10\\
 & 25 &  13 52 51.6 &  +38 39 00 &  $<$0.11   &     &       -   &      &   0.36(0.06) & 2/3 &   0.27(0.09) & 3/8\\
VCN, VCS
 & 1 &  15 15 14.5 &  +56 29 36 &  $<$0.10   &     &     -      &      &   0.10(0.03) & 2/5 &   0.10(0.03) & 3/2\\
 & 2 &  15 15 20.8 &  +56 30 43 &  $<$0.10   &     &     -      &      &   0.08(0.03) & 2/0 &   0.15(0.03) & 2/0\\
 & 3 &  15 15 26.3 &  +56 32 06 &  $<$0.10   &     &     -      &      &   0.13(0.04) & 2/3 &   0.11(0.03) & 2/1\\
 & 4 &  15 14 28.3 &  +56 10 26 &  $<$0.09   &     &     -      &      &   0.16(0.04) & 3/1 &   0.19(0.04) & 2/0\\
 & 5 &  15 14 38.7 &  +56 34 06 &   0.11(0.04) & 3/5 &      -   &      &   0.22(0.05) & 2/0 &   0.21(0.06) & 2/0\\
 & 6$^*$ &  15 16 48.0 &  +56 26 15 &   0.07(0.03) & 2/1 &  -      &      &   0.12(0.03) & 2/2 &   0.17(0.05) & 2/0\\
 & 7$^*$ &  15 16 49.1 &  +56 26 23 &   0.16(0.06) & 2/1 &  -      &      &   0.12(0.03) & 2/2 &   0.17(0.05) & 2/0\\
 & 8$^*$ &  15 16 50.7 &  +56 26 18 &   0.16(0.06) & 2/1 &   -     &      &   0.12(0.03) & 2/2 &   0.12(0.05) & 2/0\\
 & 9$^2$ &  15 15 54.3 &  +56 19 26 &  19.02(2.70) & 3/14 &  -  &      &  34.94(3.69) & 3/4 &  37.99(4.27) & 4/10\\
 & 10 &  15 14 56.7 &  +56 35 21 &   0.05(0.02) & 2/0 &     -   &      &   0.08(0.04) & 2/3 &   0.14(0.04) & 2/0\\
ZW\,IIb
 & 1$^3$ &  05 10 48.4 &  -02 40 45 &     -    &      &   0.55(0.12) & 2/0 &     -       &      &   0.62(0.21) & 3/2\\
 & 2 &  05 10 56.2 &  -02 45 41 &         -    &      &   0.38(0.13) & 2/2 &     -       &      &   0.52(0.18) & 2/0\\
\hline 
\end{tabular}
$^1$ Mrk\,314
$^2$ NGC\,5907 (see Sect.~\ref{sect:iras})
$^3$ Zw\,II\,b
\end{table*}

The surface density of sources is estimated by dividing the number of
sources counted at each flux density level by the total effective map area.
With ``effective'' map area we mean that portion of the total map area from
which such sources could have been detected. For example, in the field VCN
faint sources cannot be detected close to the bright galaxy NGC\,5907. The
maps around Mrk\,314 and ZW\,IIB were originally observed in order to study
a known object and since these bright sources are not the result of random
selection they are excluded from the source counts.

As a first approximation we have calculated the effective map areas
according to the faintest detected source. At each flux density level the
area was taken to be the sum of the areas of those maps where sources with
equal or lower flux densities were detected. The procedure is likely to
underestimate the surface density of faint sources since it ignores the
background surface brightness variations within the maps. The detection of
just one source with a flux density $S_{\nu}$ increases the corresponding
area $A(S_{\nu})$ by the the area of the whole map. On the other hand, for
statistical reasons the faintest detected source can sometimes be brighter
than the faintest theoretically detectable source but this should not lead
to significant underestimation of the areas at any flux density levels.

For 90$\mu$m maps observed with a 180$\arcsec$ raster step there are gaps
between the rasters. We have chosen not to correct for this effect but note
that it may lead to an underestimate of the source densities by some tens of
per cents i.e. the effect is comparable with the calibration uncertainties.

The cumulative source densities obtained at 90$\mu$m, 150$\mu$m and
180$\mu$m are shown as histograms in Fig.~\ref{fig:counts}. Two sets of
sources were used in deriving these curves. The first set consists of all
detections (i.e. $P>$99\%) and no confirmation was required at a different
wavelength (dotted line). The second set contains only those
sources that were confirmed by a detection at another wavelength (solid line).

The first set may contain a number of false detections. The second set gives
a conservative estimate for the true number of sources. Since the areas used
in deriving the surface densities depend also on the selection criteria
applied, the sample with the larger number of sources does not necessarily
lead to higher source density. 
The results obtained from the two sets are very similar for 150 and
180$\mu$m. At 90$\mu$m the counts based on detections confirmed at 150$\mu$m
or 180$\mu$m are, however, below the other estimates. Most of the difference
can be explained by statistical uncertainties. There are only three
confirmed 90$\mu$m sources brighter than 400\,mJy and at lower flux levels
where the number of sources is larger the estimates are in fair agreement
with each other. The change in the ratio between the two estimates as the
function of source flux is not statistically significant. Some of the
difference may be caused also by the source properties. Sources detected at
150$\mu$m are likely to be seen also at 180$\mu$m (and vice versa) while
more of the 90$\mu$m sources remain unconfirmed at the longer wavelengths.

We have derived a third set of cumulative source densities by selecting
sources based on the ratio $\rho$ between the flux density and the
background surface brightness variations (see Sect.~\ref{sect:false}). For
each flux density level all sources with $\rho>\rho_0$ were selected and no
confirmation at other wavelengths was required. The area corresponding to a
given flux density level was obtained by first calculating the local
standard deviation of the surface brightness values around each observed
position and then integrating the total area with noise below 1/$\rho_0$
times the source flux density.

The limiting value, $\rho_0=$10.5, was selected based on simulations (see
Sect.~\ref{sect:false}) which indicate that close to the detection limit the
number of false detections is still clearly less than the expected number of
true sources. In the measurements the typical background noise is 0.02\,Jy
per pixel and the given $\rho$ value corresponds to source flux density
limits 45\,mJy and 200\,mJy for C100 and C200, respectively. For higher flux
densities the number of false detections drops rapidly while for lower flux
densities more of the true sources are either rejected based on the $\rho$
criterion or are not detected at all. At the quoted flux levels the
cumulative source counts are higher than the number of the false detections
which will therefore not affect the results significantly. However,
especially at 90$\mu$m, the number of false detections can exceed the number
of undetected true sources and the source counts could be slightly
overestimated.

The cumulative source densities obtained with this third method are drawn in
Fig~\ref{fig:counts} with dashed lines. These results are based on the
integrated area of the regions where background noise is low enough for a
source with given flux density to be detectable. Since the area
determination and the source detection are based on similar criteria the
errors caused by an incorrect $\rho$--limit tend to cancel out. At the
bright end the results agree with the earlier histograms since no sources
are rejected and the corresponding areas converge towards the total area
mapped.  The differences are more pronounced at faint flux densities. When
the area corresponding to a faint flux density limit was determined based on
the faintest observed source the area was likely to be overestimated and the
source densities underestimated. When the area was determined by the local
properties of the background brightness and the sources were selected using
a related criterion the results should be more reliable. On the other hand,
the applied $\rho$ limit excludes many of the fainter (but more uncertain)
sources that were included in the previous counts and the curves cannot be
extended reliably to equally low flux densities. 
The values obtained for 150$\mu$m and 180$\mu$m below 150\,mJy are probably only indicative.

\subsection{Cumulative flux densities} \label{sect:fluxes}

\begin{figure}
\resizebox{8cm}{!}{\includegraphics{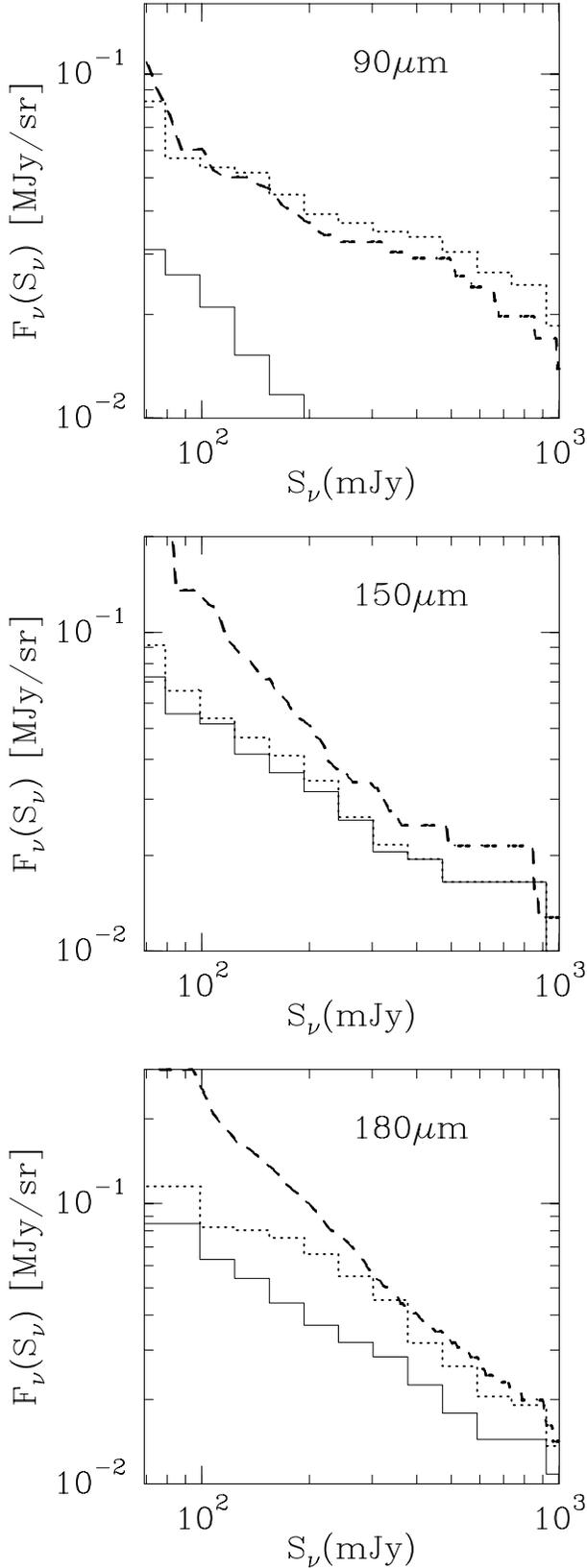}}
\caption[]{Cumulative flux densities, $F_{\nu}$, at 90$\mu$, 150 $\mu$m and
180$\mu$m in units of surface brightness. The curves correspond to samples
obtained with the same methods as in Fig.~\ref{fig:counts} }
\label{fig:fluxes}
\end{figure}

Fig.~\ref{fig:fluxes} shows the cumulative flux densities,
$F_{\nu}(S_{\nu})$, i.e. the surface brightnesses due to all sources
brighter than a given flux density $S_{\nu}$. Compared with source counts
the flux density values are more sensitive to bright sources.

Systematic calibration errors would cause the curves to be shifted
horizontally. Our simulations show that the source flux densities depend
also on the detection process. The fitted source positions can be slightly
displaced from the true positions towards a direction where the background
noise produces the highest values. Therefore it is possible that the
faintest flux densities are overestimated. The errors are, however, only
noticeable close to the detection limit and should never be more than 10\%.
Furthermore, several maps were observed with partly overlapping rasters.
The improved sampling should lead to more reliable flux densities. The largest
uncertainty is therefore in the absolute calibration itself.

\subsection{Association with known sources} \label{sect:iras}

Table~\ref{table:iras} lists IRAS Point Source Catalog (PSC) and
Faint Source Catalog (FSC) sources within the mapped regions. 
ISOPHOT sources detected within a $\sim$1$\arcmin$ radius of the IRAS
positions are also shown.

In the regions studied there are four sources with IRAS detections at
100$\mu$m. In EBL26 the PSC source at 1$^{\rm h}$17$^{\rm m}$10.6$^{\rm s}$
+2$\degr$17$\arcmin$10$\arcsec$ has been detected also with ISOPHOT. The
flux density obtained at 90$\mu$m, 2.6\,Jy, is somewhat higher than the IRAS
100$\mu$m value of 2.2\,Jy. The FSC source at 1$^{\rm h}$18$^{\rm
m}$05.0$^{\rm s}$ +1$\degr$58$\arcmin$59$\arcsec$ with 0.84\,Jy flux density
at 100$\mu$m has not been detected and the quoted 90$\mu$m upper limit is
110\,mJy. The reason is that the 90$\mu$m map was incompletely sampled and
the IRAS position lies between the observed rasters.

In VCS the IRAS source at 15$^{\rm h}$15$^{\rm m}$53.3$^{\rm s}$
+56$\degr$19$\arcmin$47$\arcsec$ is the bright galaxy NGC\,5907. The source
is extended and the ISOPHOT map is too narrow for the estimation of the
total flux density. The ISOPHOT values in Table~\ref{table:iras} are
obtained by fitting of a {\it point source} within an area with 5$\arcmin$
radius. The flux density at 90$\mu$m is slightly lower than the IRAS value.
In the case of Mrk\,314 the flux densities at 120$\mu$m are comparable to
the IRAS 100$\mu$m values. There are additionally a number of IRAS sources
with upper limits at 100$\mu$m. In all cases the ISOPHOT upper limits
derived at 90$\mu$m are much lower although in the case of maps observed
with 180$\arcsec$ raster steps the upper limits are not always accurate.

\begin{table*}
\caption[]{IRAS Faint Source Catalog and Point Source Catalog sources within
a $\sim$1$\arcmin$ radius of the positions listed in
Table~\ref{table:sources}}
\label{table:iras}
\end{table*}

In Table~\ref{table:simbad} we list objects from the Simbad
database that are within the mapped regions. The list contains all sources
identified as galaxies or quasars. Other types of objects (e.g. radio
sources) are listed only if they are close to a FIR detection.

\begin{table*}
\caption[]{
Sources from the Simbad database together with FIR detections within
$\sim$1.5$\arcmin$ radius. The table contains all galaxies and QSOs inside
the regions mapped and all other objects close to sources detected in the ISOPHOT maps}
\label{table:simbad}
\end{table*}

\section{Discussion} \label{sect:discussion}

\subsection{Comparison with other ISOPHOT source counts} \label{sect:allcounts}

At the flux level of 100\,mJy the following source densities are obtained
(dotted line in Fig.~\ref{fig:counts}): 1.4$\times$10$^5$\,sr$^{-1}$,
2.5$\times$10$^5$\,sr$^{-1}$ and 3.5$\times$10$^5$\,sr$^{-1}$ at 90\,$\mu$m,
150\,$\mu$m and 180\,$\mu$m. In Fig.\ref{fig:allcount} these results are
compared with results from other ISOPHOT projects.

At 90$\mu$m the results of Kawara et al. (\cite{kawara}) are some 30\%
higher than our source counts while the results of the ELAIS survey
(Oliver et al. \cite{ringberg}) and Linden-V{\o}rnle et al.
(\cite{linden}) are lower than our counts. The ELAIS counts based on
observations of 11.6 square degrees at 90$\mu$m, are a factor of three
below the Kawara et al. (\cite{kawara}) value.

The calibration adopted in the ELAIS project is based on DIRBE and it was
found that the PIA analysis resulted in higher surface brightnesses
(Efstathiou et al. \cite{efstathiou00}). This is consistent with our
findings. The difference between the DIRBE surface brightness values and
our data calibrated with PIA version 7.3 is $\sim$30\%
(\ref{appendix:calibration}). With DIRBE calibration our source count
points at 90$\mu$m move towards smaller flux densities (see
Fig.~\ref{fig:allcount}) and they would be in good agreement with the
ELAIS results.

At 180$\mu$m we can compare our source counts with Kawara et al.
(\cite{kawara}) and Puget et al. (\cite{puget99}) observations at 175$\mu$m.
Our counts are almost two times higher than the Kawara et al. results but
compatible with Puget et al. (\cite{puget99}).

\subsection{Comparison with galaxy models}

\begin{figure}
\resizebox{\hsize}{!}{\includegraphics{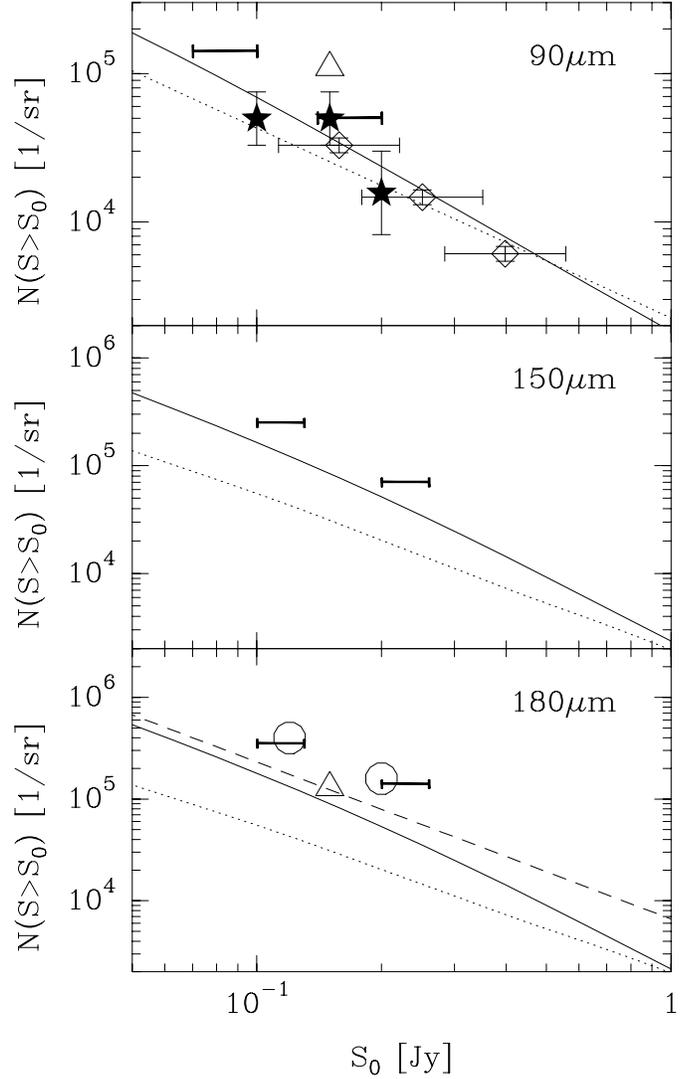}}
\caption[]{
Comparison with other ISOPHOT counts and models of galaxy evolution. Our
source counts at 100\,mJy and 200\,mJy are shown together with the results
of Puget et al. (\cite{puget99}; circles), Kawara et al. (\cite{kawara};
triangles), Oliver et al. (\cite{ringberg}; diamonds) and
Linden-V{\o}rnle (\cite{linden}; stars). We present our results as
horizontal lines that indicate the difference between the DIRBE calibration
and the adopted ISOPHOT calibration. At 90$\mu$m the DIRBE calibration
results in lower flux densities and at 150$\mu$m and 180$\mu$m in higher
flux densities. The predictions of model E of Guiderdoni et al.
\cite{guiderdoni98} are shown with solid lines and the evolutionary model of
Franceschini et al.
\cite{franceschini98} with a dashed line.
Dotted lines show predictions of no-evolution models
(90$\mu$m: Guiderdoni et al. \cite{guiderdoni98}; 150$\mu$m and
180$\mu$m: Franceschini et al. \cite{franceschini98})
}
\label{fig:allcount}
\end{figure}

At 150$\mu$m and 180$\mu$m the counts are much higher than predicted by
no-evolution models (e.g. Guiderdoni et al. \cite{guiderdoni98}); at
180\,$\mu$m the difference is a factor of five (see Fig.~\ref{fig:allcount}).

In the evolutionary model E by Guiderdoni et al. (\cite{guiderdoni98}) both
the star formation rate and the relative number of ULIRGs increase with $z$.
The model has been found to be in good agreement with extragalactic
background light measurements in both optical and FIR. Our source counts
exceed, however, the model predictions at all three wavelengths (see
Fig.~\ref{fig:allcount}).

Franceschini et al. (\cite{franceschini98}) have presented similar models
which include contributions from two galaxy populations: dust-enshrouded
formation of early-type galaxies and late-type galaxies with enhanced
star-formation at lower redshifts. The predicted source counts at 180$\mu$m
are higher than in the Guiderdoni et al. model E and the model is therefore
in better agreement with our results.

\subsection{Extragalactic background radiation}

Measurements of the CIRB in the wavelength range of the present observations
have been recently published based on DIRBE and FIRAS observations of the
COBE satellite. After the removal of interplanetary and galactic foreground
sources (Kelsall et al. \cite{kelsall}; Arendt et al. \cite{arendt}) the
level of CIRB detected by DIRBE was found to be
25$\pm$7\,nW\,m$^{-2}$\,sr$^{-1}$ at 140\,$\mu$m and
14$\pm$3\,nW\,m$^{-2}$\,sr$^{-1}$ at 240\,$\mu$m (Hauser et al.
\cite{hauser}). These numbers correspond to 1.1\,MJy\,sr$^{-1}$. According
to Fig.~\ref{fig:fluxes} the sources found in this study represent
some 5\% of the CIRB, the exact number depending on which source counts are
applied.

Fixsen et al. (\cite{fixsen98}) have reported 
results based on three different methods used to subtract the galactic
foreground emission. From the analytical representation of the average
spectrum the surface brightness values at 90\,$\mu$m, 150\,$\mu$m and
180$\mu$m are 0.13\,MJy\,sr$^{-1}$, 0.67\,MJy\,sr$^{-1}$ and
0.82\,MJy\,sr$^{-1}$, respectively. Adopting these values our sources
would contribute already a significant fraction of the CIRB, especially at
90$\mu$m where $\ga$20\% of the CIRB could be attributed to the detected
sources. For sources brighter than 100\,mJy the corresponding fraction at
150\,$\mu$m is slightly below and at 180$\mu$m slightly above 10\%.
Recalibrating the 140$\mu$m and 240$\mu$m DIRBE observations using the
results of the cross-calibration between FIRAS and DIRBE (Fixsen et al.
\cite{fixsen97}) would, naturally, move the DIRBE points closer to the FIRAS
values (Fixsen et al. \cite{fixsen98}). As discussed in
\ref{appendix:calibration}, our calibration appears to be closer to the
FIRAS than the DIRBE scale.

\subsection{Comparison with galaxy spectra} \label{sect:followup}

\begin{figure}
\resizebox{\hsize}{!}{\includegraphics{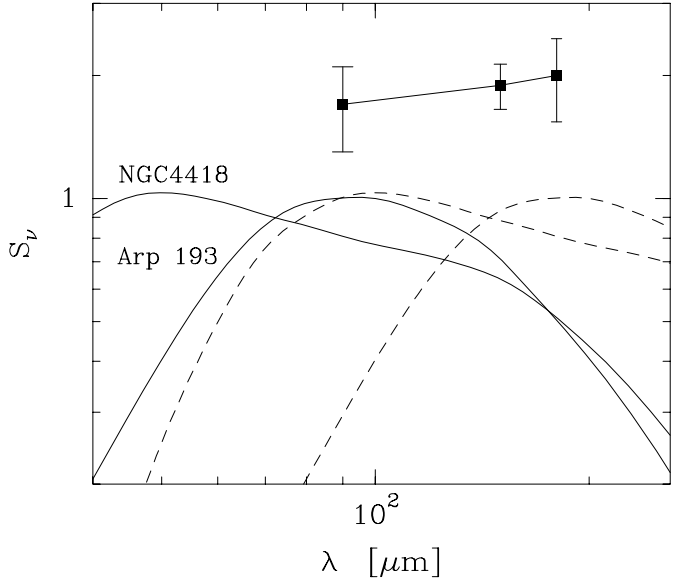}}
\caption[]{
The average of the source spectra shown in
Figs.~\ref{fig:90det_vc}-\ref{fig:90det_ngp} and the two-temperature model
spectra of Lisenfeld et al. (\cite{lisenfeld}) for the luminous infrared
galaxies Arp\,193 and NGC\,4418. Dashed lines show the spectra of the two
galaxies shifted to $z$=1.0. The flux density scale is arbitrary}
\label{fig:lisenfeld}
\end{figure}

In Fig.~\ref{fig:lisenfeld} we compare the average of the source spectra
presented in Figs.~\ref{fig:90det_vc}-\ref{fig:90det_ngp} with the spectra
of the galaxies Arp\,193 and NGC\,4418. In the sample of luminous infrared
galaxies presented by Lisenfeld et al. (\cite{lisenfeld}) Arp\,193
has the lowest and NGC\.4418 the highest estimated dust temperature.

In the rest frame the SEDs of luminous infrared galaxies reach maxima
between 60$\mu$m and 100$\mu$m (Silva et al. \cite{silva}; Devriendt et al.
\cite{devriendt}; Lisenfeld et al. \cite{lisenfeld}). The spectra
of our FIR sources are relatively flat in the observed wavelength range and
peak typically above 90$\mu$m. This is consistent with most sources being at
redshifts $0.5\la$$z$$\la$1.

The emission maximum of normal spiral galaxies is also close to 100$\mu$m
(e.g. Silva et al. \cite{silva}). However, in the case of a spiral galaxy a
FIR detection at the level of 0.1\,Jy would correspond to a visual magnitude
brighter than 16 and the optical counterpart should be visible. The lack of
visual counterparts indicates that most of our sources are likely to be
distant luminous infrared galaxies.

We have conducted a follow-up study in the VCN region. A
$\sim$2$\times$2$\arcmin$ field surrounding the positions of the FIR sources
VCN 1 and 2 in Table~\ref{table:sources} has been observed in $V$- and $I$-bands
(images provided by J.C. Cuillandre, CFHT) and in $K-$ and $J$-bands (images
taken by P. V\"ais\"anen).

\begin{figure}
\resizebox{\hsize}{!}{\includegraphics{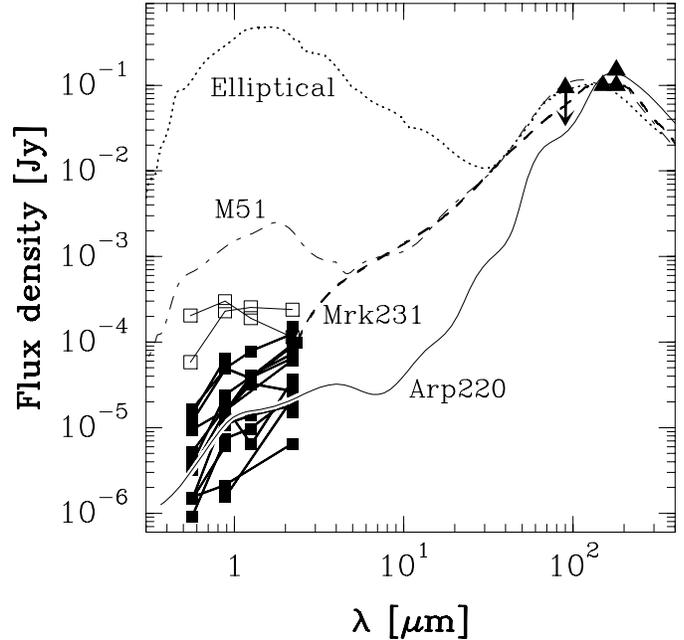}}
\caption[]{
Flux densities of sources with $I$--$K>3.0$ (filled squares) together with
the flux densities of the two ISO detections (at 90$\mu$m only an upper
limit) in a field within the VCN region. The spectra of the brightest
sources with arbitrary $I$--$K$ within 50$\arcsec$ of the FIR source
positions are shown as open squares. The spectral energy distributions of
two luminous infrared galaxies, Arp\,220 (solid line) and Mrk\,231 (dashed
line), normalized to the 150$\mu$m observations, are drawn for $z\sim$1.0
(Ivison et al. \cite{ivison98}). For comparison, template spectra of an
elliptical galaxy (dotted line) and of the spiral galaxy M51 (dash-dotted)
are shown for $z=0$ (Silva et al. \cite{silva})}
\label{fig:fluxsed}
\end{figure}

Our photometry reveals a number of red sources with $I-K\ga3$. These are
potential candidates for being luminous infrared galaxies (LIRG) and could
be the counterparts of our FIR detections. In Fig.~\ref{fig:fluxsed} we show
the flux densities of these sources together with the ISO detections. In the
same figure the spectral energy distributions of two luminous infrared
galaxies, Arp\,220 and Mrk\,231, are drawn as they would be seen at redshift
$1.0$ (Ivison et al. \cite{ivison98}). The figure shows that, in principle,
any one of the red NIR sources could be responsible for our ISO FIR
detections. An unambiguous identification is not possible. Normal
elliptical or spiral galaxies, as shown in the figure, are excluded.

\section{Conclusions} \label{sect:conclusions}

We have searched for FIR point sources in raster maps observed with the
ISOPHOT C100 and C200 detectors at wavelengths between 90$\mu$m and
180$\mu$m. The total area covered is $\sim$1.5\,square degrees. Most of the
FIR sources detected are presumably IR galaxies which, due to the negative
$K$-correction, can be observed at redshifts $z>1$. A comparison of the
SEDs of sources detected at 90, 150, and 180$\mu$m with cirrus spectra shows
that for most sources an explanation in terms of cirrus knots can be
excluded. Based on the number counts of the sources we can conclude:

\begin{itemize}
\item  We have found 55 FIR sources that, due to the multi-wavelength
confirmation, correspond to detections with high confidence level
($\ga$4$\sigma$)
\item We have derived a FIR source density of $\sim$60 sources per $\sq\degr$ at
100\,mJy level
\item The source density is much higher than predicted by no-evolution
galaxy models; at 180$\mu$m the excess is close to a factor of five
\item The source counts are in agreement with models where the star formation
rate and the relative number of ULIGs increases strongly with $z$, e.g. the
counts are slightly higher than predicted by model E of Guiderdoni et al.
(\cite{guiderdoni98})
\item
At 150$\mu$m and 180$\mu$m the combined flux of detected sources
accounts for $\sim$10\% of the CIRB intensity as derived from the COBE FIRAS
data; at 90$\mu$m the fraction is over 20\%

\end{itemize}

\begin{acknowledgements}
We thank J.C. Cuillandre for providing the $V-$ and $I-$images of the VCN
region and P.V\"ais\"anen for the NIR images. This study was supported by
the Academy of Finland Grant no. 1011055. ISOPHOT and the Data Centre at
MPIA, Heidelberg, are funded by the Deutsches Zentrum f\"ur Luft- und
Raumfahrt and the Max-Planck-Gesellschaft.
\end{acknowledgements}

\appendix

\section{Calibration comparisons} \label{appendix:calibration}

The results presented in this paper are based on the calibration performed
with the onboard FCS. This has been estimated to be better than 30\% for
extended sources brighter than 4\,MJy\,sr$^{-1}$ (Klaas et al.
\cite{klaas98}). Because of the very low surface brightness and the
correspondingly low FCS power, it is interesting to compare the ISO
calibration scale with the corresponding DIRBE values.

For the comparison of absolute surface brightnesses we took the DIRBE ZSMA
({\em Zodi-Subtracted Mission Average Maps}) data around each ISO field.
Zodiacal light was added to the DIRBE values according to the model given by
Leinert et al. (\cite{leinert98}) using the $\lambda-\lambda_{\sun}$ and the
ecliptic latitude of the ISO observations. We could also have used the
Weekly Averaged Sky Map data observed during those weeks when the solar
elongation is similar as in the ISO observations. Because of the somewhat
lower noise we chose to use the the ZSMA data. Both DIRBE and ISO were
colour corrected assuming a $\nu^2 B_{\nu}$ spectrum and $T_{\rm
dust}$=18\,K (Arendt et al. \cite{arendt}; Schlegel et al.
\cite{schlegel98}). DIRBE data were interpolated to the wavelength of the
ISO observations. The average ISO flux density weighted with the DIRBE beam
was compared with the DIRBE value. The surface brightness ratios obtained
are shown in Table~\ref{table:dirbe1} in column 4. The main uncertainties
are due to the large DIRBE beam and the large noise of the DIRBE 140\,$\mu$m
and 240\,$\mu$m observations in these faint surface brightness regions. The
given error estimates correspond to the total dispersion in the surface
brightness values. These estimates are probably more realistic than those
obtained by combining the error estimates calculated for the mean surface
brightness values.

\begin{table}
\caption[]{
Comparison of the DIRBE and ISO flux density scales for surface
brightnesses. The columns are: (1) the field name, (2) wavelength of ISO
observations, (3) mean surface brightness of the ISO map, (4) the ratio of
the DIRBE and ISO absolute surface brightnesses. For column 4 the numbers
given in parentheses give the dispersions in the DIRBE surface brightness
values interpolated to the wavelength of the ISO observations (see text). 
Note that for EBL22 we have chosen to correct the calibration according to
existing absolute photometry measurements
}
\label{table:dirbe1}
\begin{tabular}{lccc}
\hline
Field   &  $\lambda$
        & $<S>$  
	&  $S_{\rm DIRBE}/S_{\rm ISO}$  
	\\
        & ($\mu$m) &  (MJy\,sr$^{-1}$) &               \\
(1) & (2) & (3) & (4)  \\
\hline

EBL26
& 180 & 6.0     &  1.44(0.30)      \\
& 150 & 7.9     &  1.40(0.29)      \\
& 90  & 15.8    &  0.79(0.05)      \\

NGPS
& 180  & 3.1    &  1.29(0.17)  \\
& 150  & 3.7    &  1.24(0.15)  \\
& 90   & 5.7    &  0.65(0.06)  \\

NGPN
& 180  & 1.8     &  1.67(0.07) \\
& 180$^1$ & 2.4  &  1.28(0.03) \\
& 150  & 2.5     &  1.43(0.07) \\
& 90   & 5.5     &  0.57(0.02) \\

NGPN \& NGPS
& 180  & 2.2    &  1.44(0.23)  \\
& 150  & 3.3    &  1.31(0.18)  \\
& 90   & 5.6    &  0.62(0.06)  \\

EBL22 
& 180  & 2.3     &  2.25(0.09) [1.21(0.05)]      \\
&      &         & {\it 1.73(0.07)}       \\
& 150  & 2.0     &  2.00(0.07) [1.24(0.05)]        \\
&      &         & {\it 1.69(0.06)}       \\
& 90   & 0.8     &  0.80(0.02)          \\
&      &         & {\it 0.85}            \\

ZW II  
& 180  & 12.6    &  1.28(0.03)     \\
& 120  &  9.6    &  1.59(0.04)     \\

\hline
\end{tabular}
\vspace{0.2cm}

$^1$ {\it the larger 180$\mu$m map} \\
{\it [\,] using default responsivities} \\
{\it italic entries}: corrected according to absolute photometry measurements
\end{table}

At 90$\mu$m the DIRBE values are $\sim$30\% lower than the ISO values. On
the other hand, there is an indication that the ISO C200 values are
$\sim$30-40\% lower than the DIRBE ones.

At 140$\mu$m and at 240$\mu$m the relative error estimates given for ZSMA
surface brightness values exceed in many cases 50\%. When the temperature
was fixed to 18\,K the interpolated values at 150\,$\mu$m and 180\,$\mu$m
are mostly determined by the selected temperature and the 100\,$\mu$m DIRBE
values which have significantly smaller error estimates. An increase of
$T_{\rm dust}$ by 1\,K increases $S_{\rm DIRBE}/S_{\rm ISO}$ at 90\,$\mu$m
by $\sim$1\% and decreases it at 150$\mu$m and 180$\mu$m by about 10\%. On
the other hand, if the exponent in $\nu^{\alpha}$ is decreased from 2.0 to
1.0 the discrepancy between ISO and DIRBE scales would increase at 180$\mu$m
by $\sim$20\%. If all sources of uncertainty are taken into account our
results do not indicate a difference between the ISO and DIRBE flux density
scales exceeding $\sim$30\%.

In the case of EBL22 there was an unusually large difference, by a factor of
$\sim$0.6, between the values obtained with default responsivities and those
calibrated with the FCS measurements. The fact that the FCS heating power
was clearly below the range for which calibration tables exist may have
contributed to the large discrepancy. Comparison with DIRBE supports the
higher surface brightness values obtained with the default responsivities
(see Table\ref{table:dirbe1}). Absolute photometry at one position in the
field gives values closer to the default calibration. Since the FCS
calibration is more reliable in the case of the absolute photometry we have
re-scaled the maps to agree with the absolute photometry. For all the other
maps the differences between the FCS and the default responsivities remained
below 30\%.

\section{Details of the point source detection} \label{appendix:detection}

\subsection{Flat fielding}  \label{appendix:flatfield}

Normally the flat fielding was done by calculating the ratios between each
detector pixel and the average of other measurements within a small area
around it. Linear fits provided the flat fielding correction factors as 
function of the surface brightness.

In some cases (e.g. in VCN) it was found advantageous to perform the flat
fielding partially together with the source fitting. This was done by adding
free parameters to scale independently the measurements made with different
detector pixels. This introduces to the fit three additional parameters in
the case of C200 maps and eight parameters for the C100 maps. Determining
the flat fielding (multiplicative factor only) locally makes it possible to
automatically correct for some detector drifts. The method is useful when
the detector drifts are slow compared with the time needed to observe the
region surrounding a point source.

\subsection{The fitting procedure} \label{appendix:fitting}

The point source detection was performed in two steps using the surface
brightness values and their errors, one value for each detector pixel in
each raster position. In the first step measurements more than
0.7$\sigma_{\rm bg}$ above the local background level were flagged as point
source candidates. The background level and its dispersion, $\sigma_{\rm
bg}$, were estimated from other measurements within a radius which was
typically three times the size of the detector pixel.

In the second step a model of a point source and a background was fitted to
each region surrounding the candidate positions. The background was assumed
to be constant, since in most cases the gradient of the background was small.
The free parameters of the fit were the source flux density, the two
coordinates of the source position, and the background surface brightness.

The contribution of a point source to the measured flux density at different
map positions was computed using the footprint matrices of PIA. Footprint
matrices provide the fraction of the flux detected by each detector pixel.
Some 70\% of the flux from a point source located at the centre of a pixel
will be detected by this one pixel. The detector beams are approximately
gaussian in their central part but have more extended wings. The neighboring
pixels will therefore receive slightly more flux than predicted by the
gaussian approximation, about 7\% or less for source distances exceeding one
pixel step (46.0$\arcsec$ for C100 and 92$\arcsec$ for C200).

The radius of the region used in the fitting procedure was typically
$\sim$2.2$\arcmin$ for maps observed with the detector C100 and
$\sim$3.7$\arcmin$ for maps done with C200. These radii are large enough to
make the contribution from the point source small at the edge of the
fitting region.

\subsection{Completeness and false detections in simulations} \label{appendix:false}

\begin{figure}
\resizebox{\hsize}{!}{\includegraphics{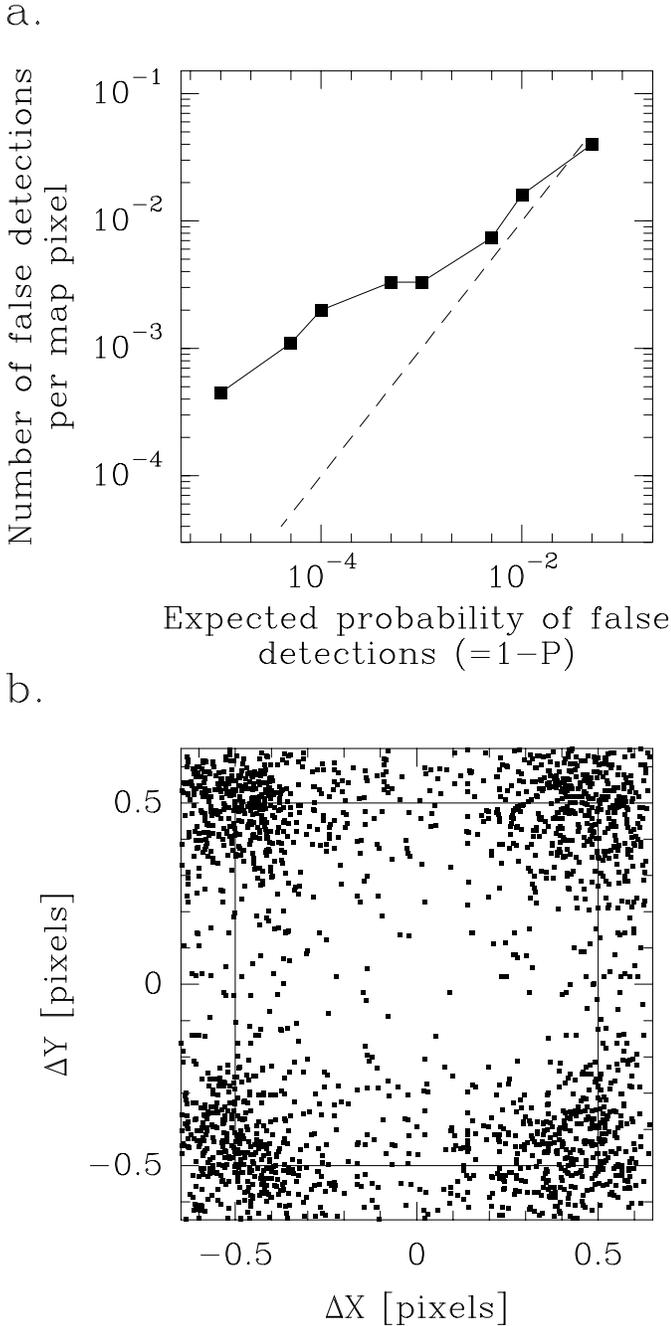}}
\caption[]{
{\bf a} The number of false detections per map pixel in simulated
measurements as a function of the expected number 1-$P$. The confidence
level $P$ is used to discard uncertain detections. At high confidence levels
the actual number of false detections exceeds the expected number (dashed
line). {\bf b} Positions of false detections inside a map pixel for
non-overlapping rasters. Similar distribution exists for very weak sources }
\label{fig:simu}
\end{figure}

The completeness of the point source detection and the probability of false
detections were studied with simulated maps with raster step equal to the
detector size i.e. without any redundancy. We simulated the dependence of
faint sources detection on the background noise level. The criterion given
above was used for selecting the candidate pixels with point source
contribution, the probability limit, $P>$99\%, was used to discard
uncertain detections. Different values of the ratio 
\begin{equation}
\rho = \frac{S_{\nu}}{\sigma_{\rm bg}}
\end{equation}
between source flux density, $S_{\nu}$, given in Jy and the background rms
noise, $\sigma_{\rm bg}$, given in units of Jy were also
considered. The detection rate is about 90\% for $\rho
=7.8$ while for values close to $\rho=$5.2 the detection rate drops below
50\%. The number of false detections as the function of the probability
limit $P$ was also studied. The number of false detections is, as expected,
linear with respect to $P$. However, the number of false sources was found
to exceed the expected number of 1-$P$ false detections per map pixel (see
Fig.~\ref{fig:simu}a). This was taken into account when determining the
parameters of source detection. One would obviously like to have criteria
where the number of false detections is equal to the unknown number of
undetected real sources.

The flux densities of false detections were found to lie between
$S_{\nu}\approx 3.3\sigma_{\rm bg}$ and below $S_{\nu}\approx 7.0\sigma_{\rm
bg}$. The lower limit is set partly by the initial selection of candidate
pixels that are 0.7$\sigma_{\rm bg}$ above the background noise. The spatial
distribution of the false detections is shown in Fig.~\ref{fig:simu}b. They
are concentrated close to the edges and the corners of the pixels. The
source density varies roughly as $r^2$ where $r$ is the distance from the
pixel centre. This does not, however, help in eliminating false detections
since the spatial distribution is similar for very faint true sources. The
bias is present only close to the detection limit.

\subsection{Influence of imperfect flat fielding and de-glitching}
\label{appendix:fferror}

The simulations described above, that were used in determining suitable
detection limits, do not take into account some effects that may affect the
source counts.
For example, imperfect flat fielding or de-glitching will lead to higher
surface brightness values in some detector pixels and may increase the
probability of classifying some random variations as point sources. The
problem is potentially more serious for the C100 detector where the
deviation of one detector pixel has less influence on the overall noise of
the surface brightness estimated for the region. In those cases where the
local flat fielding coefficients were estimated together with the source
parameters the quality of the flat fielding is less important. In order to
estimate its effect in other cases we simulated C100 measurements consisting
first of normally distributed noise and then scaled upwards the values of
one detector pixel. The number of false detections did not, however, depend
strongly on the imperfections of the flat fielding and was in fact smaller
for the tested range where one detector pixel deviated less than 2$\sigma$.
If more than one pixel deviates from the mean the effect is further reduced.
We conclude that imperfect flat fielding will not cause significant errors
in the source counts.

\section{Methods in the examination of SRD data} \label{appendix:visual}

The main procedures used in the source extraction were based on AAP
(Astrophysical Applications Data) data products reduced with PIA. The
corresponding SRD (Signal per Ramp Data) files were used for visual
inspection of the data. 
Glitches are visible at the SRD level as a sudden high signal value followed
by gradually decreasing tail or, in the case of several glitches close in
time, as unusually large noise. The number of ramps per raster position
varies in our data from 6 to over 40. In the case of the C100 detector,
which is more affected by the glitches, the number of ramps was always
sufficient to see the characteristic features of the glitches.

The SRD data close to the potential sources were inspected visually in order
to check whether the detections were possibly due to glitches and not to
real point sources. The raster position and the detector pixel closest to
the fitted source position was determined and with the selected raster
position in the centre the SRD data for five consecutive raster position and
for all detector pixels were plotted (see Fig.~\ref{fig:ebl26srd}). 
Both the variations in the signal level between raster positions and between
the different detector pixels were examined by eye and the measurement
closest to the source position was checked for signs of glitches. A quality
flag between 0 and 4 was given for the source and these are shown in
Table~\ref{table:sources} as flag ``q''. The intended scale is:

\begin{itemize}
\item 0: the median signal is clearly affected by glitches and/or the noise is
clearly too high for reliable determination of the median signal
\item 1: there are clear glitches that may have affected the median signal
and/or large noise makes the determination of the median uncertain 
\item 2: no clear signs of a source, no significant glitches
\item 3: a possible source
\item 4: a clear detection
\end{itemize}

The difference in the signal levels between different raster positions and
between different detector pixels were also used in the scrutinization.
Therefore, even if the signals were seriously affected by glitches the
source could be classified in e.g. class 3 provided that the true signal
level could be estimated to be high enough.

The classification is, of course, subjective. It is also strongly biased
towards sources that happen to lie in the centre of some detector pixel and
are therefore visible only in one measurement. Furthermore, if the source is
situated between two raster lines the previously described SRD vs. time
plots do not show all data relevant for the classification.

The main benefit from the eyeball inspection is that we can recognize
potentially false detections (classes 0 and 1) that are due to detector
glitches. There are, however, only a couple of such possibly false
detections in our source sample. This shows that most false detections
were avoided in our source detection procedure.

The source detections were tested also by applying to the SRD data methods
reminiscent of the procedures used in the ELAIS project (Surace et al.
\cite{surace99}).

For each source we determined the raster position and the detector pixel,
$p_{\rm 0}$ closest to the potential source. Using the signal values $s_{\rm
i}^{\rm p}$ from all detector pixels $p$ from five consecutive raster
positions ($i$=-2\ldots 2) centered on the selected position we calculated
the following values:

\begin{enumerate}
\item $s_{\rm 0}^{\rm p_{\rm 0}}/s_{\rm 0}^{\rm p}, p\neq p_{\rm 0}$
subtracted by the median of the corresponding values at the five raster positions
\item same as (1.) but instead of the median we use a prediction from 2nd degree
fit to the five raster positions
\item $s_{\rm 0}$ subtracted by the median of $s_{\rm i}^{\rm p_{\rm 0}}$,
$i$=-2\ldots2
\item same as previous but using prediction from a 2nd degree fit instead of
the median
\item $s_{\rm 0}^{\rm p_{\rm 0}}$ divided by the median of other pixels in
the current raster position subtracted by the median of the corresponding
values in all five raster positions
\item same as previous but using a 2nd degree fit instead of the median
of five raster positions
\end{enumerate}

In the case of C100 we calculated therefore altogether 20 values and in the
case of C200 10 values. The number of values above 1.5$\sigma$ level was
counted and is shown in Table~\ref{table:sources} as flag ``$c$''. It must be
emphasized that the method is applied only to the SRD data closest to the
previously determined source positions and it is therefore most sensitive to
sources that were close to the centre of some detector pixel. Thus a low
value of ``$c$'' does not disqualify a source detection obtained with our
detection procedure.

\end{document}


\special{landscape}

\setcounter{table}{4}

\begin{table*}
\caption[]{Galaxies and quasars in the mapped regions from the Simbad
database. The list contains also other sources when they are located close
to our FIR detections. The columns are: (1) the name of the source, (2)-(3)
coordinates, (4) the classification of the source, (5)-(12) data based on
ISOPHOT observations: positional distances from Simbad source and our
estimates for the FIR flux densities. In case no FIR sources were detected
close to the Simbad positions estimated upper limits for the flux densities
are given
}

\begin{tabular}{lcclrrrrrrrr}
\multicolumn{4}{c}{Simbad} & \multicolumn{8}{c}{ISOPHOT} \\
name &  RA & DEC & type & 
 \multicolumn{2}{c}{90$\mu$m} &\multicolumn{2}{c}{120$\mu$m} &
 \multicolumn{2}{c}{150$\mu$m} &\multicolumn{2}{c}{180$\mu$m} \\

  &  (2000.0) & (2000.0) &   & 
  $\Delta$ ($\arcsec$) & $S$ (mJy) & $\Delta$ ($\arcsec$) & $S$ (mJy) & 
  $\Delta$ ($\arcsec$) & $S$ (mJy) & $\Delta$ ($\arcsec$) & $S$ (mJy) \\
 (1) & (2) & (3) & (4) & (5) & (6) & (7) & (8) & (9) & (10) & (11) & (12) \\
\hline
 \multicolumn{12}{l}{EBL22} \\
APMBGC 478-116+053 & 02 23 51.4 &  -25 49 17 &    G   &    &   $<$110   &    &     &    &   $<$270   &  75 &   367(99) \\
\hline
 \multicolumn{12}{l}{EBL26} \\
Z 0114.6+0201 & 01 17 10.0 &  +02 16 47 &    G   &  57 &   2644(1035)  &    &     &  17 &   1823(332)  &  26 &   1684(373) \\
LSBG 0114+0204 & 01 16 48.9 &  +02 20 40 &  LSB   &    &   $<$170   &    &     &    &   $<$1130   &    &   $<$1000  \\
PC 0116+0126A & 01 19 25.3 &  +01 42 24 &  EmO   &    &   $<$190   &    &     &    &   $<$380   &    &   $<$280  \\
PC 0116+0126B & 01 19 25.4 &  +01 42 16 &  EmO   &    &   $<$190   &    &     &    &   $<$340   &    &   $<$250  \\
PC 0116+0123 & 01 19 27.0 &  +01 39 22 &  EmO   &    &   $<$170   &    &     &    &   $<$340   &    &   $<$220  \\
BHB98 J011910.0+013128 & 01 19 10.0 &  +01 31 27 &  QSO   &  43 &   144(51)  &    &     &    &   $<$220   &  38 &   163(48) \\
TXS 0117+011 & 01 19 47.3 &  +01 24 59 &  Rad   &    &   $<$150   &    &     &    &   $<$300   &  92 &   214(60) \\
\hline
 \multicolumn{12}{l}{Mrk 314} \\
UGC 12329 & 23 02 59.3 &  +16 36 18 &    G   &    &     &  2 &   1317(174)  &    &     &  22 &   907(168) \\
\hline
 \multicolumn{12}{l}{NGPN} \\
KUG 1344+399 & 13 46 27.7 &  +39 40 40 &    G   &    &   $<$140   &    &     &    &   $<$500   &    &   $<$460  \\
FIRST J134613.2+393855 & 13 46 13.3 &  +39 38 55 &  Rad   &    &   $<$160   &    &     &    &   $<$480   &  34 &   606(95) \\
FIRST J134612.7+393848 & 13 46 12.7 &  +39 38 48 &  Rad   &    &   $<$160   &    &     &    &   $<$480   &  44 &   606(95) \\
ACO 1789$^1$ & 13 46 09.0 &  +39 39 00 &  ClG   &    &   $<$150   &    &     &    &   $<$510   &  70 &   606(95) \\
FIRST J134609.3+395459$^1$ & 13 46 09.3 &  +39 54 59 &  Rad   &    &   $<$280   &    &     &  54 &   174(58)  &    &   $<$200  \\
FIRST J134424.3+400209 & 13 44 24.4 &  +40 02 09 &  Rad   &    &   $<$110   &    &     &    &   $<$190   &  91 &   220(79) \\

Mrk 460$^1$ & 13 44 12.6 &  +40 00 07 &    G   &    &   $<$80   &    &     &    &   $<$170   &  92 &   220(79) \\
FIRST J134230.1+402542 & 13 42 30.1 &  +40 25 42 &  Rad   &    &   $<$100   &    &     &    &   $<$210   &  80 &   214(70) \\
FIRST J134236.3+402813 & 13 42 36.4 &  +40 28 13 &  Rad   &  60 &   162(35)  &    &     &  91 &   226(55)  &  81 &   286(74) \\
FIRST J134152.9+402749 & 13 41 52.9 &  +40 27 49 &  Rad   &    &   $<$70   &    &     &    &   $<$230   &  79 &   233(92) \\

\end{tabular}
\end{table*}

\begin{table*}
\begin{tabular}{lcclrrrrrrrr}
\multicolumn{4}{c}{Simbad} & \multicolumn{8}{c}{ISOPHOT} \\
name &  Ra & Dec & type & 
 \multicolumn{2}{c}{90$\mu$m} &\multicolumn{2}{c}{120$\mu$m} &
 \multicolumn{2}{c}{150$\mu$m} &\multicolumn{2}{c}{180$\mu$m} \\

  &  (2000.0) & (2000.0) &   & 
  $\Delta$ ($\arcsec$) & $S$ (mJy) & $\Delta$ ($\arcsec$) & $S$ (mJy) & 
  $\Delta$ ($\arcsec$) & $S$ (mJy) & $\Delta$ ($\arcsec$) & $S$ (mJy) \\
 (1) & (2) & (3) & (4) & (5) & (6) & (7) & (8) & (9) & (10) & (11) & (12) \\

\hline
 \multicolumn{12}{l}{NGPN (the larger 180$\mu$m map)} \\
ZwCl 1341+4022 & 13 43 24.0 &  +40 07 00 &  ClG   &    &     &    &     &    &     &  29 &   137(28) \\
FIRST J134418.9+402259 & 13 44 19.0 &  +40 22 59 &  Rad   &    &     &    &     &    &     &  76 &   405(97) \\
FIRST J134446.1+402952 & 13 44 46.2 &  +40 29 52 &  Rad   &    &     &    &     &    &     &  33 &   174(63) \\
FIRST J134454.9+403215 & 13 44 54.9 &  +40 32 15 &  Rad   &    &     &    &     &    &     &  47 &   187(76) \\
KUG 1341+407 & 13 44 05.9 &  +40 31 23 &    G   &    &     &    &     &    &     &  20 &   367(64) \\
FIRST J134346.1+402906 & 13 43 46.1 &  +40 29 06 &  Rad   &    &     &    &     &    &     &  61 &   298(75) \\
NPM1G +40.0332 & 13 42 07.6 &  +40 15 32 &    G   &    &     &    &     &    &     &    &   $<$170  \\
RX J1343.0+4032 & 13 42 59.0 &  +40 32 30 &    X   &    &     &    &     &    &     &  48 &   265(75) \\
FIRST J134252.9+403201 & 13 42 53.0 &  +40 32 01 &  QSO   &    &     &    &     &    &     &    &   $<$260  \\
RX J1340.6+4018$^1$ & 13 40 33.4 &  +40 17 48 &  ClG   &    &     &    &     &    &     &    &   $<$260  \\
FIRST J134242.3+405242 & 13 42 42.3 &  +40 52 42 &  Rad   &    &     &    &     &    &     &  75 &   220(54) \\
FIRST J134153.7+404601 & 13 41 53.7 &  +40 46 01 &  Rad   &    &     &    &     &    &     &  27 &   335(65) \\
MCG+07-28-047 & 13 40 02.9 &  +40 25 15 &    G   &    &     &    &     &    &     &  42 &   366(72) \\
FIRST J134214.6+405221 & 13 42 14.6 &  +40 52 21 &  Rad   &    &     &    &     &    &     &  61 &   208(84) \\
Z 1339.5+4108$^2$ & 13 41 42.2 &  +40 52 28 &    G   &    &     &    &     &    &     &  12 &   958(72) \\

\hline 

 \multicolumn{12}{l}{NGPS} \\
B3 1345+398 & 13 47 13.5 &  +39 35 57 &  Rad   &  66 &   194(40)  &    &     &  77 &   206(85)  &  74 &   205(85) \\
KK98a 134521.9+393726$^1$ & 13 47 30.9 &  +39 22 29 &    G   &    &   $<$90   &    &     &    &   $<$150   &    &   $<$310  \\
FIRST J134826.3+392652 & 13 48 26.3 &  +39 26 52 &  Rad   &    &   $<$140   &    &     &  72 &   225(41)  &  20 &   139(56) \\
FIRST J134921.5+390926 & 13 49 21.5 &  +39 09 26 &  Rad   &    &   $<$120   &    &     &  64 &   173(74)  &    &   $<$240  \\
B3 1348+392 & 13 50 32.4 &  +38 59 22 &  QSO   &    &   $<$130   &    &     &  74 &   206(80)  &  81 &   255(94) \\
NPM1G +39.0331 & 13 50 46.0 &  +38 49 06 &    G   &  33 &   183(44)  &    &     &    &   $<$250   &    &   $<$230  \\

FIRST J135207.4+384732 & 13 52 07.5 &  +38 47 32 &  Rad   &    &   $<$110   &    &     &    &   $<$240   &  42 &   223(73) \\
UGC 8793 & 13 52 38.0 &  +38 41 11 &    G   &    &   $<$190   &    &     &    &   $<$290   &    &   $<$290  \\

\end{tabular}
\end{table*}

\begin{table*}
\begin{tabular}{lcclrrrrrrrr}
\multicolumn{4}{c}{Simbad} & \multicolumn{8}{c}{ISOPHOT} \\
name &  Ra & Dec & type & 
 \multicolumn{2}{c}{90$\mu$m} &\multicolumn{2}{c}{120$\mu$m} &
 \multicolumn{2}{c}{150$\mu$m} &\multicolumn{2}{c}{180$\mu$m} \\

  &  (2000.0) & (2000.0) &   & 
  $\Delta$ ($\arcsec$) & $S$ (mJy) & $\Delta$ ($\arcsec$) & $S$ (mJy) & 
  $\Delta$ ($\arcsec$) & $S$ (mJy) & $\Delta$ ($\arcsec$) & $S$ (mJy) \\
 (1) & (2) & (3) & (4) & (5) & (6) & (7) & (8) & (9) & (10) & (11) & (12) \\

\hline
 \multicolumn{12}{l}{VCN} \\
CRSS J1515.1+5628 & 15 15 10.1 &  +56 28 34 &  QSO   &    &   $<$80   &    &     &  65 &   100(27)  &  80 &   99(30) \\
RX J1515.1+5628 & 15 15 10.2 &  +56 28 34 &    X   &    &   $<$80   &    &     &  64 &   100(27)  &  78 &   99(30) \\
GHCG 152$^1$ & 15 16 24.0 &  +56 24 00 &  GrG   &    &   $<$60   &    &     &    &   $<$2850   &    &   $<$2160  \\
KUG 1513+566$^1$ & 15 14 48.1 &  +56 27 16 &    G   &    &   $<$90   &    &     &    &   $<$240   &    &   $<$260  \\
RX J1515.3+5630 & 15 15 21.0 &  +56 30 48 &    X   &  92 &   35(15)  &    &     &  28 &   78(33)  &  7 &   154(34) \\
RX J1514.5+5633 & 15 14 33.0 &  +56 33 00 &    X   &    &   $<$90   &    &     &  76 &   216(49)  &  73 &   214(58) \\
RX J1514.8+5634 & 15 14 51.0 &  +56 34 29 &    X   &  57 &   51(17)  &    &     &  77 &   84(36)  &  78 &   142(42) \\

\hline
 \multicolumn{12}{l}{VCS} \\
Z 1514.6+5630 & 15 15 54.0 &  +56 19 45 &    G   &  8 &   19022(2703)  &    &     &  51 &   34945(3692)  &  5 &   37990(4272) \\
V84 99 & 15 16 06.0 &  +56 19 00 &  GrG   &  72 &   9571(1981)  &    &     &  86 &   34945(3692)  &    &   $<$16540  \\
GHCG 152 & 15 16 24.0 &  +56 24 00 &  GrG   &    &   $<$120   &    &     &    &   $<$5040   &    &   $<$3150  \\
CRSS J1517.0+5623 & 15 17 03.6 &  +56 23 37 &  QSO   &    &   $<$120   &    &     &    &   $<$210   &    &   $<$230  \\

\hline
 \multicolumn{12}{l}{ZW} \\
ZW II 33 & 05 10 48.1 &  -02 40 53 &    G   &    &     &  17 &   547(117)  &    &     &  31 &   618(208) \\

\hline
\end{tabular}

$^1$ Source close to the border of the mapped area

$^2$ Galaxy Z~1339.5+4108 does not exist in our list of source detections.
It lies at the edge of the large 180$\mu$m map i.e. it has been observed at
only one wavelength. We did include into the source list those sources that
could be detected in maps obtained by dividing the data of this 180$\mu$m
map at the SRD level into two halves. At the SRD level each measurement of a
map position contained just 6 signal values i.e. after the division each of
the two maps had 3 signals per position. At the position of the galaxy five
of the six signals were high but the last one was below the background
level. The galaxy was therefore clearly detected in the first map but the
confidence obtained from the second remained below our detection threshold.
The quoted flux density estimate is based on all of the 180$\mu$m data.

\end{table*}